\documentclass[a4paper,11pt]{article}
\usepackage{jinstpub}
\usepackage{hyperref}
\usepackage{orcidlink}
\usepackage{comment}
\usepackage{caption}
\usepackage[toc,page]{appendix}
\usepackage{csquotes}
\usepackage{subcaption}
\usepackage{amsmath}
\usepackage{graphicx}
\usepackage{makecell}

\title{Optical characterization of wavelength-shifting and scintillating-wavelength-shifting fibers
}

\author[a,1]{W.~Bae\orcidlink{0000-0002-7646-7577}\note{Corresponding author.},}
\author[a]{J.~Cesar\orcidlink{0000-0001-6644-0023},}
\author[a]{K.~Chen\orcidlink{0009-0008-5642-7624},}
\author[a]{J.~Cho\orcidlink{0009-0004-7530-7231},}
\author[a]{D.~Du\orcidlink{0009-0003-5668-2282},}
\author[a]{J.~Edgar\orcidlink{0009-0005-6003-8689},}
\author[a]{L.~Earthman\orcidlink{0009-0004-2040-953X},}
\author[b]{O.M.~Falana\orcidlink{0000-0003-0158-1997},}
\author[a]{M.~Gajda\orcidlink{0009-0002-2147-0848},}
\author[b]{C.~Hurlbut,}
\author[b]{M.~Jackson,}
\author[a]{K.~Lang\orcidlink{0000-0003-1269-7223},}
\author[a]{C.~Lee\orcidlink{0009-0003-5915-3642},}
\author[a]{J.Y.~Lee\orcidlink{0009-0008-4809-3775},}
\author[a]{E.~Liang\orcidlink{0009-0008-6995-3842},}
\author[a]{J.~Liu\orcidlink{0009-0009-4037-2179},}
\author[b]{C.~Maxwell}
\author[a]{C.~Murthy\orcidlink{0000-0001-9044-7946},}
\author[a]{D.~Myers\orcidlink{0000-0001-8402-7240},}
\author[a]{S.~Nguyen\orcidlink{0009-0009-9622-6429},}
\author[b]{T.~O’Brien,}
\author[a]{M.~Proga\orcidlink{0000-0002-0303-5159},}
\author[a]{T.~Rodriguez,}
\author[a]{S.~Syed\orcidlink{0009-0000-6163-5534},}
\author[a]{M.~Zalikha\orcidlink{0009-0002-7045-6022},}
\author[a]{J.~Zey\orcidlink{0009-0008-3005-6642}}

\affiliation[a]{University of Texas at Austin, Department of Physics,
1 University Station, Austin, TX 78712-0264, USA}
\emailAdd{wonseokb@utexas.edu}

\affiliation[b]{Eljen Technology, 1300 W.\ Broadway, Sweetwater, TX 79556, USA}

\abstract{
We report results of optical characterizations of new wavelength-shifting and scintillating-wavelength-shifting fibers EJ-182 and EJ-160 from Eljen Technology and compare them to the wavelength-shifting fiber BCF-91A from Saint-Gobain.

The wavelength-dependence of attenuation was derived from spectral measurements confirming that the long attenuation length increases with wavelength, while short attenuation effects become less significant at longer wavelengths.
The impact of the environmental refractive index was studied by immersing the EJ-160II fiber in water. Immersing the fiber in water reduced the overall light output and suppressed the short attenuation component, which can be explained by reduced light-collection efficiency due to the smaller refractive-index contrast between the fiber cladding and the surrounding medium.}


\begin{document}
\maketitle
\flushbottom

%==================================================================================
%==================================================================================
\section{Introduction}

Plastic fibers, particularly those with scintillating (Sci) or wavelength-shifting (WLS) properties, or combined scintillating-wavelength-shifting (Sci-WLS), enable efficient light collection and transport by converting shorter wavelengths scintillation light to longer wavelengths that may be better matched to the spectral response of photodetectors. 
These features led to their widespread use in particle and nuclear physics experiments, including MINOS~\cite{MINOS-Michael:2008bc, Avvakumov:2005ww}, NOvA~\cite{Ayres:2004js}, and T2K ND280~\cite{T2K-ND280-NIMA}.
%, and the AMoRE-II muon veto system~\cite{AMoRE-NIMA}. 
In recent years, WLS fibers coated with tetraphenylbutadiene (TPB) have also been adapted for the collection of the scintillation light in liquid-argon in detectors such as GERDA~\cite{Ackermann:2012xja} and LEGEND-200~\cite{LEGEND:2025insdet}.

This work stems from a partnership between a group at the University of Texas at Austin and Eljen Technology~\cite{Eljen2} to develop new fibers that would not only be competitive with those available on the market, but would be better optimized for the needs of future particle physics experiments, including LEGEND-1000~\cite{LEGEND:2021bnm}. 
In this work, we report the results of the testing of three such fibers, WLS fiber EJ-182I, and Sci-WLS fibers EJ-160I and EJ-160II, and compare their performance with that of the WLS fiber BCF-91A from Saint-Gobain~\cite{Saint-Gobain2}, now Luxium Solutions~\cite{Luxium-Solutions}.

%==================================================================================
%==================================================================================
\section{Fiber samples}

\begin{figure}[h!]
    \setlength{\abovecaptionskip}{5pt}  
    \centering
    \includegraphics[width=.495\textwidth, height=.37\textwidth]{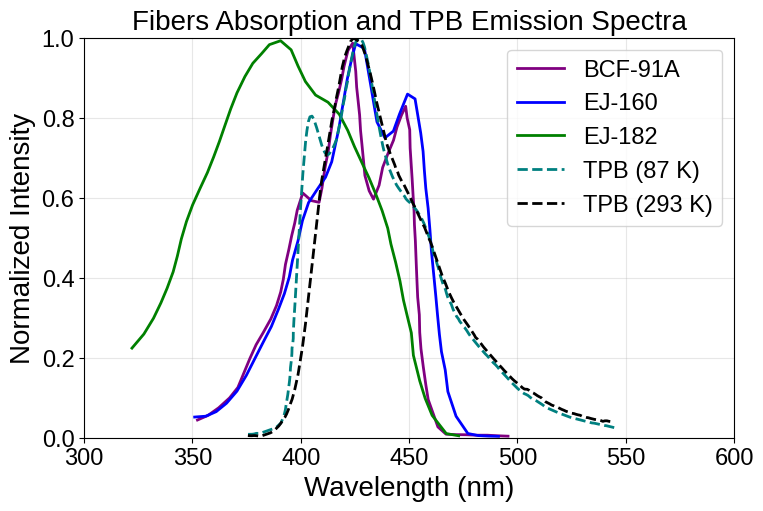}
    \includegraphics[width=.495\textwidth, height=.37\textwidth]{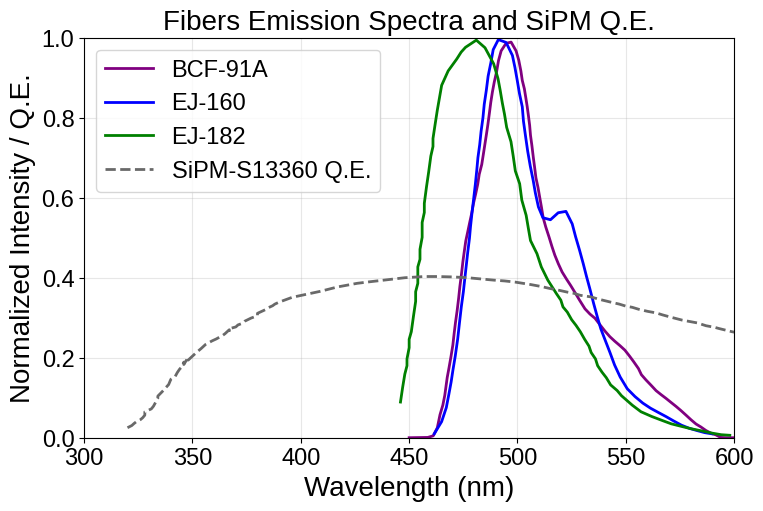}

    \caption{Absorption (left) and emission (right) spectra of the WLS fibers (manufacturers' data). 
    For reference, we include the emission spectra of tetraphenylbutadiene (TPB) from ~\cite{TPB-Leonhardt-JINST-2024}, which is often used for shifting scintillation light of liquid argon or liquid xenon, and the quantum efficiency of Silicon Photomultiplier (SiPM) S13360 from Hamamatsu Photonics~\cite{hamamatsu}. 
    All fibers and TPB spectra are normalized to their respective maxima.}
    \label{fig: fiber emission spectrum1}
\end{figure}

Table~\ref{table:wls_fibers1} and Figure~\ref{fig: fiber emission spectrum1} summarize the main physical and optical properties of the four fibers tested. 
The Saint-Gobain fiber BCF-91A type was previously used in the GERDA~\cite{Ackermann:2012xja} and LEGEND-200~\cite{LEGEND:2025insdet} experiments. The test samples were supplied to us by a group at the Technical University of Munich. 
The newly-developed EJ-160 fibers from Eljen Technology are available in two variants which have different fluor mixtures. The two variants are denoted EJ-160I and EJ-160II. Additionally, the EJ-182I fiber from Eljen Technology was also tested.
All fibers studied here have a 1~mm square cross-section (Table~\ref{table:wls_fibers1}), matching the baseline fiber size adopted for the LEGEND-1000 design~\cite{LEGEND:2021bnm}.

\begin{figure}[h!]
\centering
\includegraphics[width=.98\textwidth]{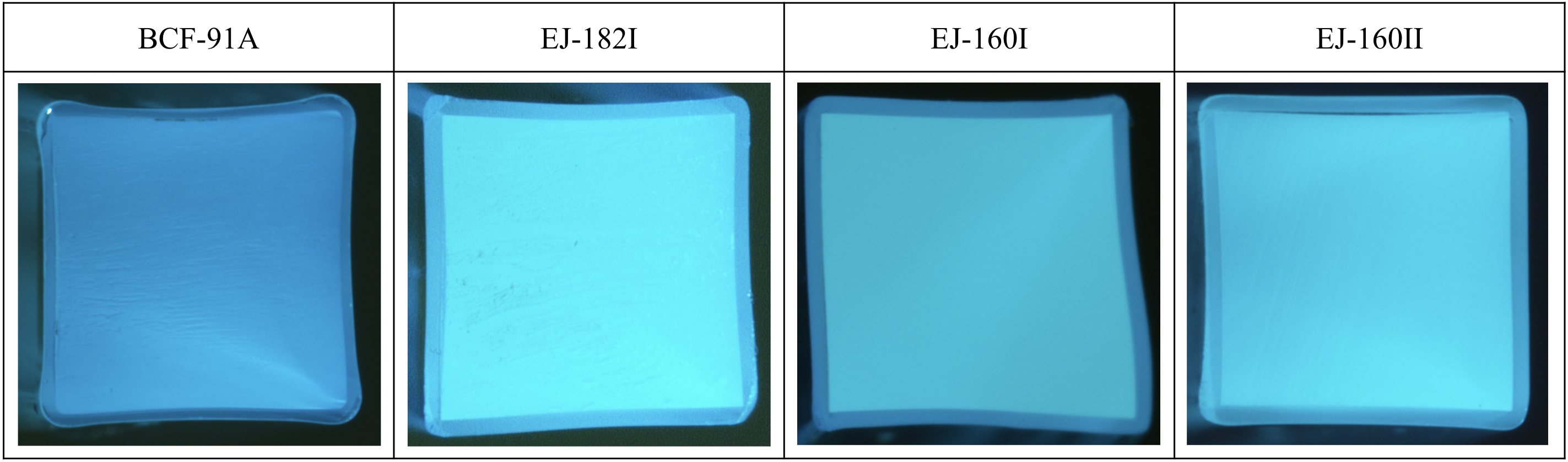}
\caption{\raggedright
Pictures of diamond fly-cut cross sections of the four tested fibers. 
These images were captured under a microscope with external illumination 
to highlight the core/cladding boundaries.}
\label{fig: microscope inspection}
\end{figure}

Figure~\ref{fig: microscope inspection} shows images of diamond fly-cut cross sections of fibers captured with a %Nikon SMZ1500 
microscope~\cite{Nikon-microscpoe}. The tested BCF-91A fiber has a single cladding of approximately 0.03\,mm thickness, while the Eljen Technology fibers each have a cladding layer of 0.04\,mm thickness. 

%==================================================================================

\begin{table}[h!]
\small
\centering
{\renewcommand{\arraystretch}{1.2}
\begin{tabular}{|p{3.5cm}|c|c|c|}
    \hline
    \multicolumn{1}{|c|}{\textbf{Feature}} 
        & \textbf{BCF-91A} 
        & \textbf{EJ-182} 
        & \textbf{EJ-160} \\
    \hline
    Variant 
        & standard 
        & EJ-182I 
        & \makecell{EJ-160I\\EJ-160II} \\
    \hline
    Type 
        & wavelength-shifting 
        & wavelength-shifting 
        & scintillating-wavelength-shifting\\
    \hline
    Cross-section 
        & 1~mm square & 1~mm square & 1~mm square \\
    \hline
    Cladding 
        & single\textsuperscript{(*)}  
        & single\textsuperscript{(*)}  
        & single\textsuperscript{(*)} \\
    \hline
    Core material  
        & polystyrene & polystyrene & polystyrene \\
    \hline
    Cladding material 
        & PMMA & PMMA & PMMA \\
    \hline
    \makecell[{{p{3.5cm}}}]{Refractive index \\ (core / cladding)}
        & 1.60/1.49 
        & 1.59/1.49 
        & 1.59/1.49 \\
    \hline
    Cladding thickness 
        & 0.03\,mm\textsuperscript{(*)} 
        & 0.04\,mm\textsuperscript{(*)} 
        & 0.04\,mm\textsuperscript{(*)} \\
    \hline
\end{tabular}
} % arraystretch scope end

\caption{Properties of tested fibers. PMMA is polymethyl methacrylate. 
\textsuperscript{(*)} The thicknesses of claddings were measured by our microscope.}
\label{table:wls_fibers1}
\end{table}

%==================================================================================
%==================================================================================
%==================================================================================
%==================================================================================

%==================================================================================
%==================================================================================
%==================================================================================
%==================================================================================

%==================================================================================
%==================================================================================

\section{Measurement setup}

\begin{figure}[h!]
        \centering
        \includegraphics[width=0.48\textwidth, height=0.48\textwidth]{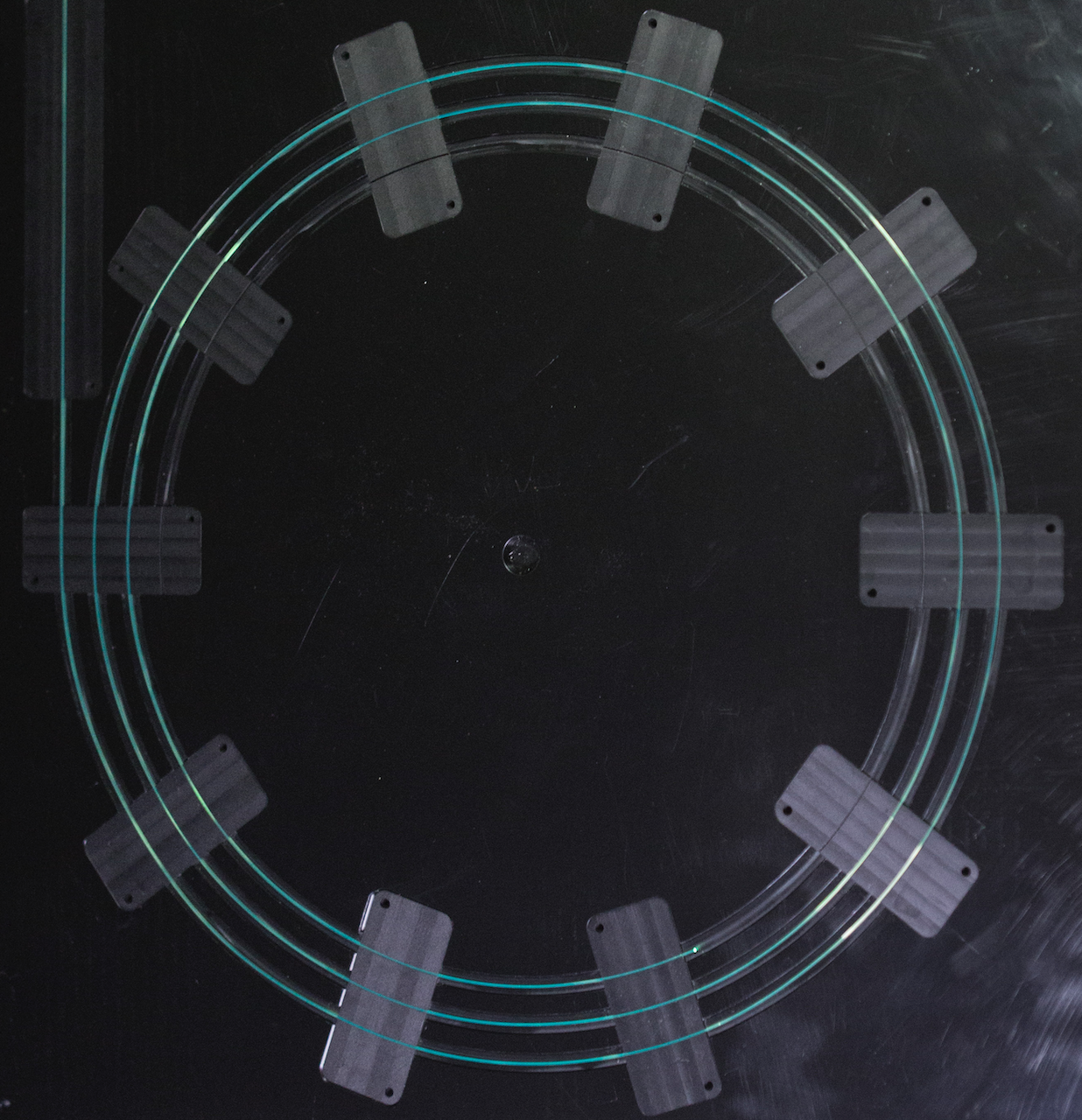}
        \hspace{0.01\textwidth}
        \includegraphics[width=0.48\textwidth, height=0.48\textwidth]{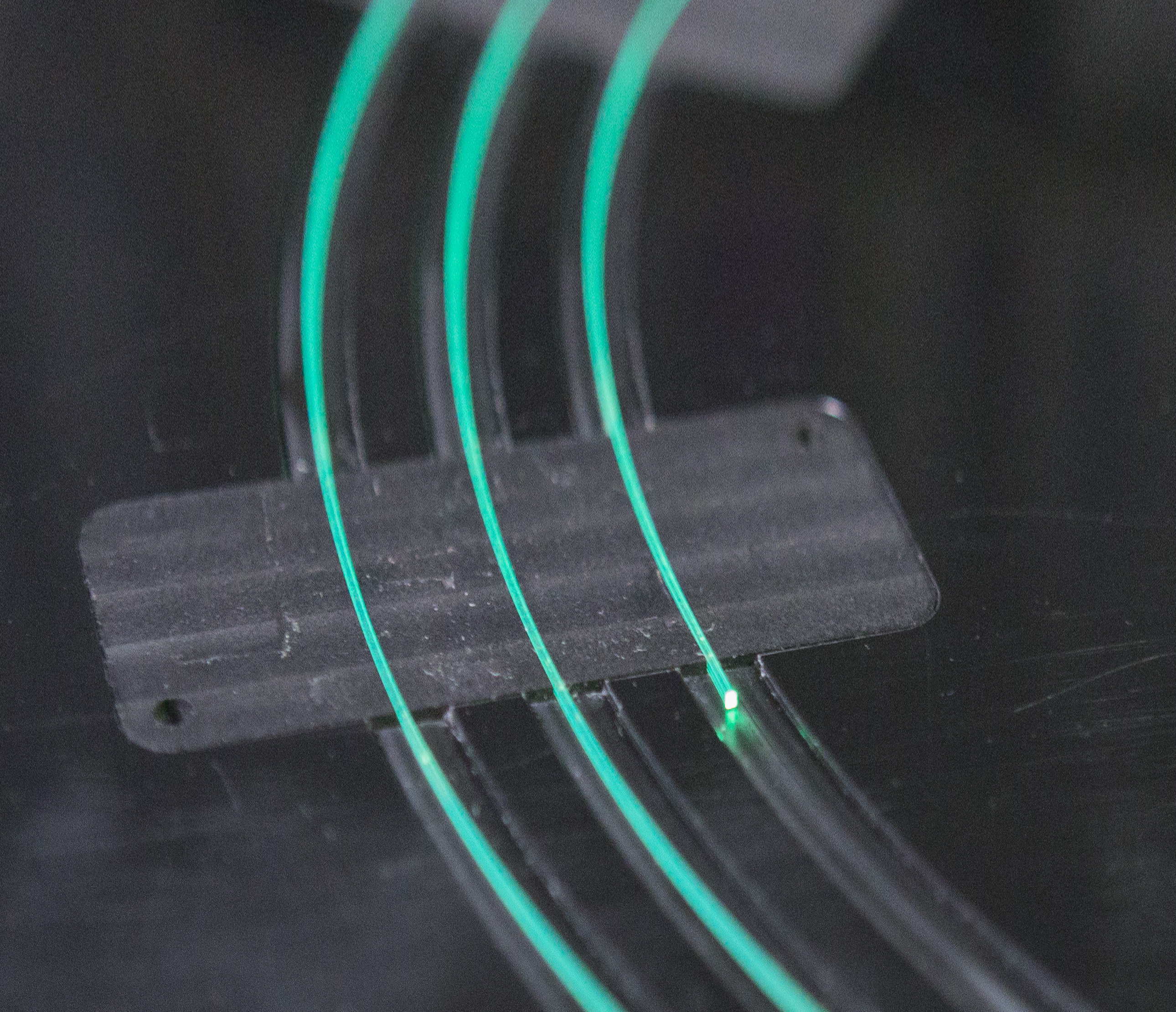}
        \caption{\raggedright 
        Photographs of the setup used in measurements. 
        Left: a fiber arranged in a spiral groove.
        Right:  A close-up view of the left picture.
        }
        \label{fig: optical bench}
        \end{figure}

The fibers were approximately 3.0\,m long. For in-air measurements, they were placed in a spiral groove.
The diameter of the groove gradually changed from 45 to 60\,cm, as shown in Figure~\ref{fig: optical bench}, ensuring a negligible transmission loss  due to bending~\cite{Fiber-bending}. 
An end of the fiber was coupled to the %Ocean Optics USB-4000 
spectrophotometer~\cite{Ocean-Optics} that was used to record the transmitted spectra. 
The instrument operates in the near-UV/visible range (approximately $350$--$800\,\mathrm{nm}$)
with a wavelength precision of about $0.21\,\mathrm{nm}$.
The opposite end of the fiber was diamond-fly-cut, polished, and left uncoated, so that light exiting that end was not collected in this characterization. In practical applications, the far end is often coupled to a photosensor or a reflective termination to maximize light collection; this increases the total collected light output but would not change the intrinsic optical properties of the fibers.
Each spectrum was the average of 10 consecutive intrinsic measurements of the spectrophotometer, each integrated over 10\,ms, and provided stable spectral profiles.

\begin{figure}[h!]
        \centering

        \includegraphics[height=0.4\textwidth]{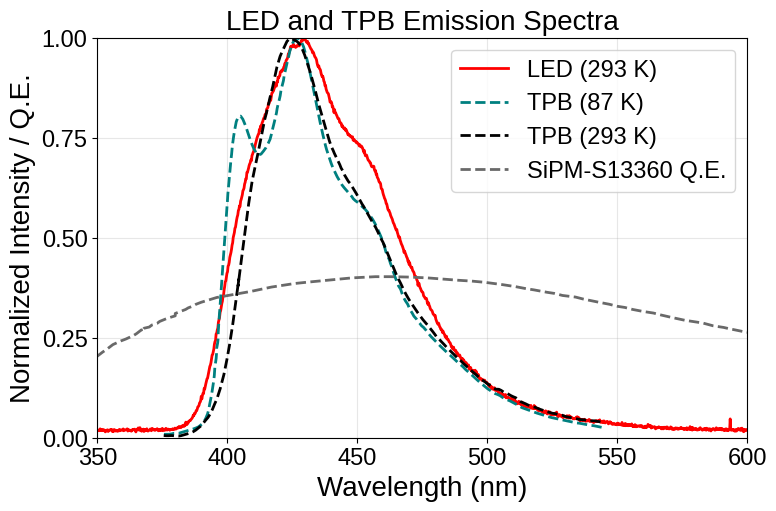}

        \caption{LED emission spectrum, measured with our spectrophotometer. For reference, the emission spectrum of TPB~\cite{TPB-Leonhardt-JINST-2024} and the quantum efficiency of a Hamamatsu SiPM S13360~\cite{hamamatsu} are also shown.}

        \label{fig: LED emission spectrum}
        \end{figure}

The bottom plate with the groove housed the fiber, while the cover plate was equipped with 20 holes for fitting an LED to illuminate a fiber. The holes were arranged to approximately uniformly span a distance ranging from 0.124\,m to 2.944\,m to the spectrophotometer.
A blue LED L200CUB500~\cite{LEDtronics} %biased to 3.8\,V, 
was selected for illuminating fibers.
The emission spectrum of this LED closely matches the TPB emission spectrum, a popular molecular wavelength shifter~\cite{Segreto2015, Araujo2022}, as shown in Figure~\ref{fig: LED emission spectrum}.
All measurements were performed in the dark at room temperature (about 21$^\circ$C) and the residual background spectrum was subtracted from the transmitted LED spectrum.

%==================================================================================
%==================================================================================
%==================================================================================
%==================================================================================
%
%==================================================================================
%==================================================================================
%==================================================================================
%==================================================================================

% \clearpage

%==================================================================================
%==================================================================================
\section{Optical characterization of fibers}

\subsection{Emission spectra}

\begin{figure}[h!]
\centering
\begin{tabular}{cc}
\includegraphics[width=.49\textwidth]{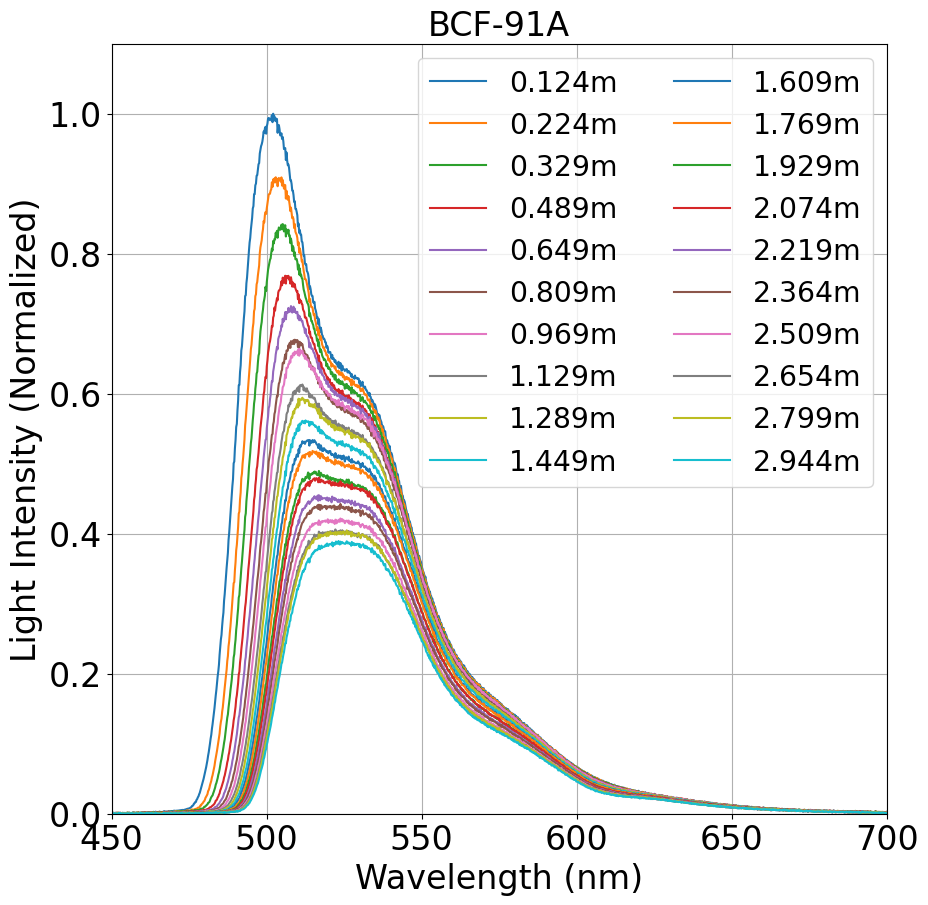} &
\includegraphics[width=.49\textwidth]{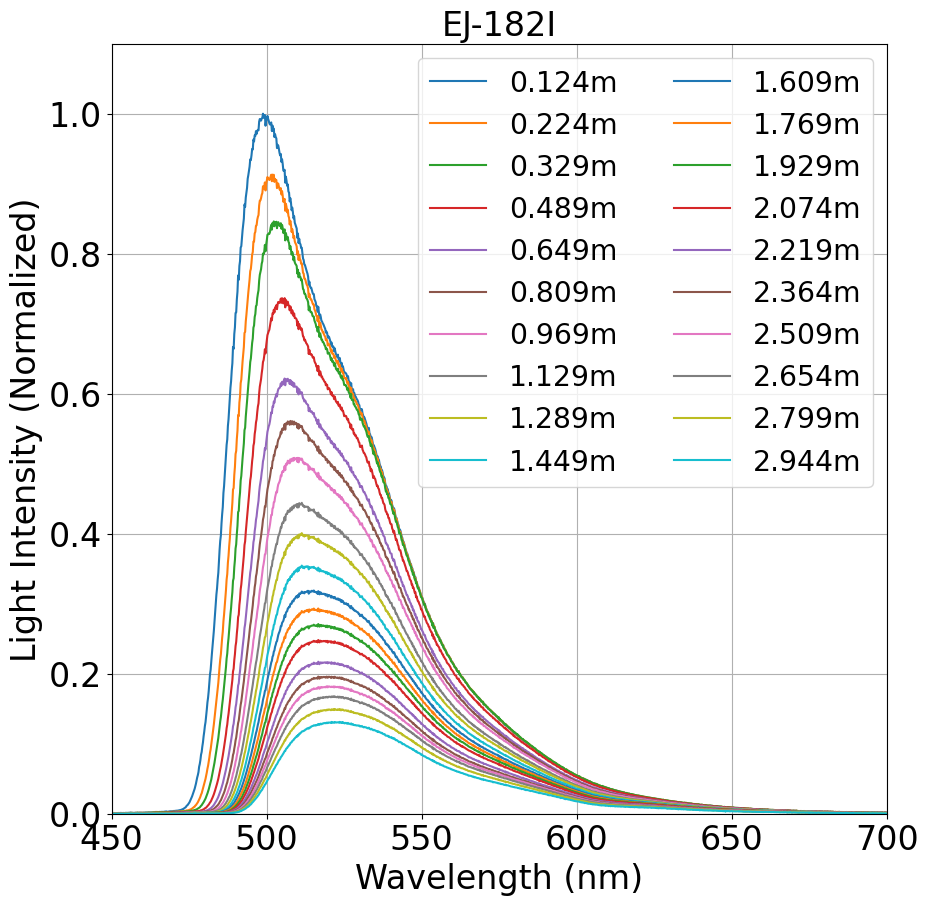} \\
\includegraphics[width=.49\textwidth]{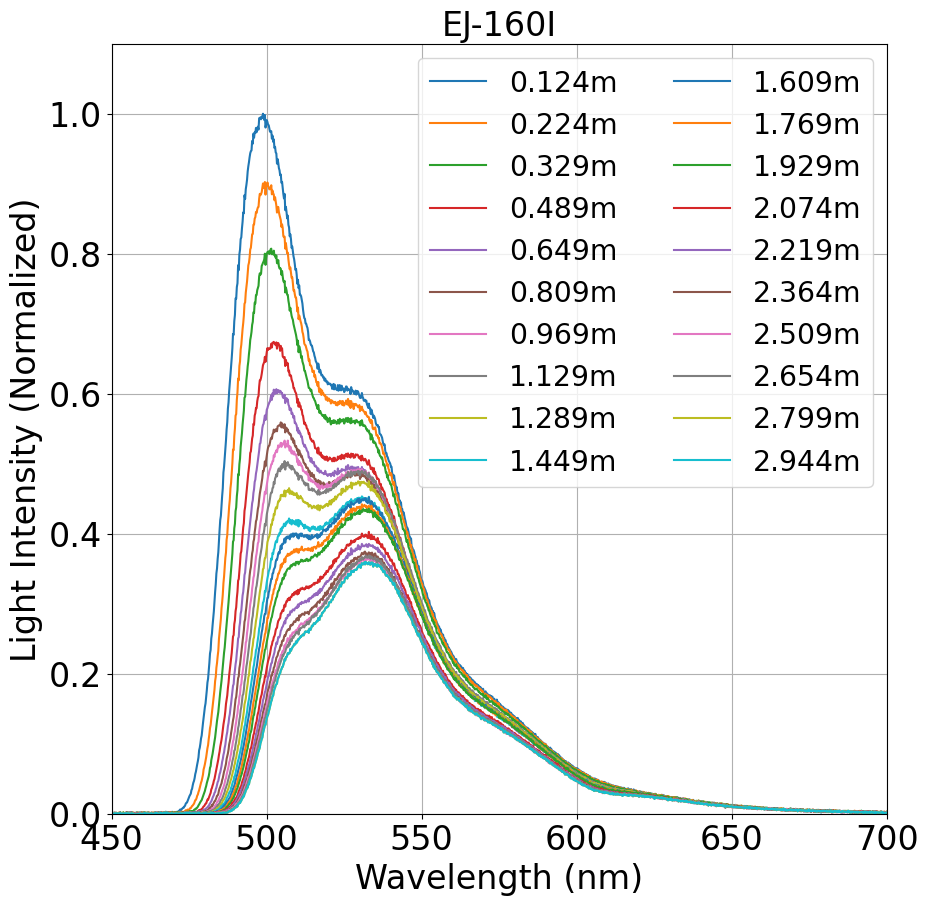} &
\includegraphics[width=.49\textwidth]{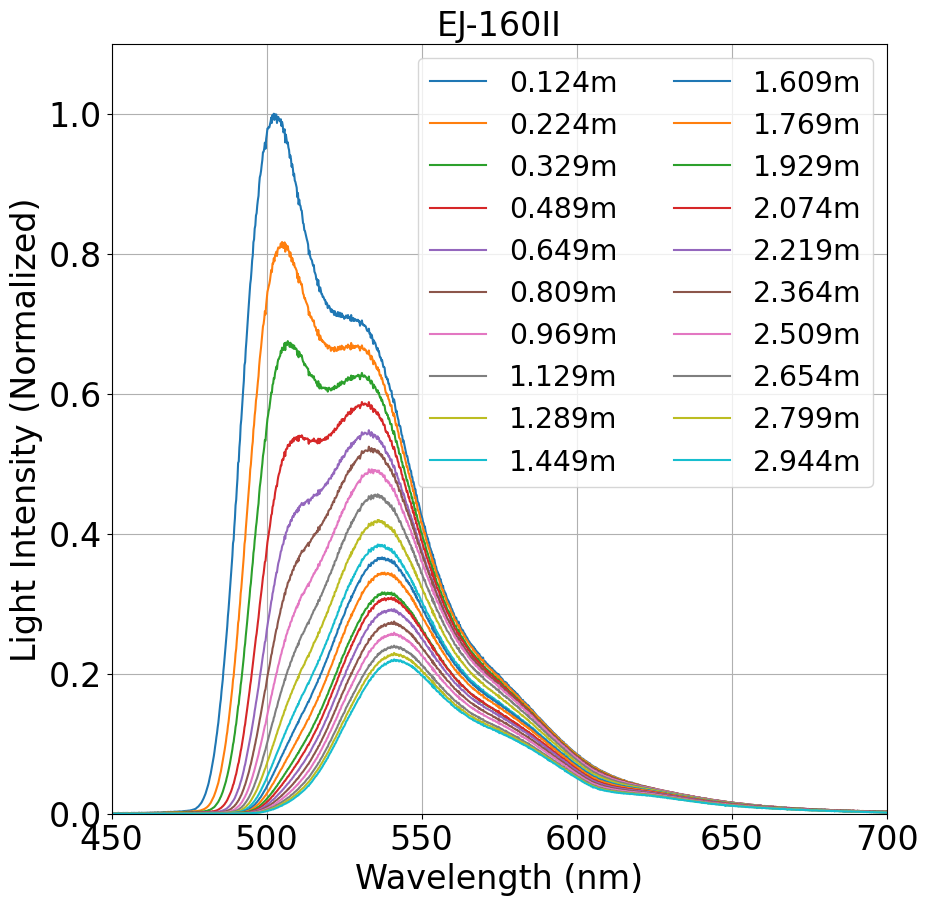}
\end{tabular}
\captionsetup{justification=centering}
\caption{Normalized emission spectra measured at 20 distances for the four tested fibers:
BCF-91A (top left), EJ-182I (top right), EJ-160I (bottom left), and EJ-160II (bottom right).}
\label{fig: emission spectrum of 4 green WLS fibers}
\end{figure}

Figure~\ref{fig: emission spectrum of 4 green WLS fibers} show the emission spectra recorded by the spectrophotometer for four tested fibers BCF-91A, EJ-182I, EJ-160I, and EJ-160II measured at 20 light propagation distances (defined as the distance between the LED illumination point and the spectrophotometer input).
The distances ranged from 0.124\,m to 2.944\,m. The spectra are normalized to unity at the maximum of the spectrum obtained with the LED placed in the hole that gives the shortest distance of 0.124\,m to the spectrometer.
At each distance, the fiber was illuminated from the side through the PMMA-based cladding of negligible absorption.

%==================================================================================

%==================================================================================

%\clearpage

%==================================================================================
%==================================================================================
\subsection{Attenuation lengths and light intensity}

\begin{figure}[h!]
    \centering
    % 1st line
    \includegraphics[width=0.495\textwidth]
    {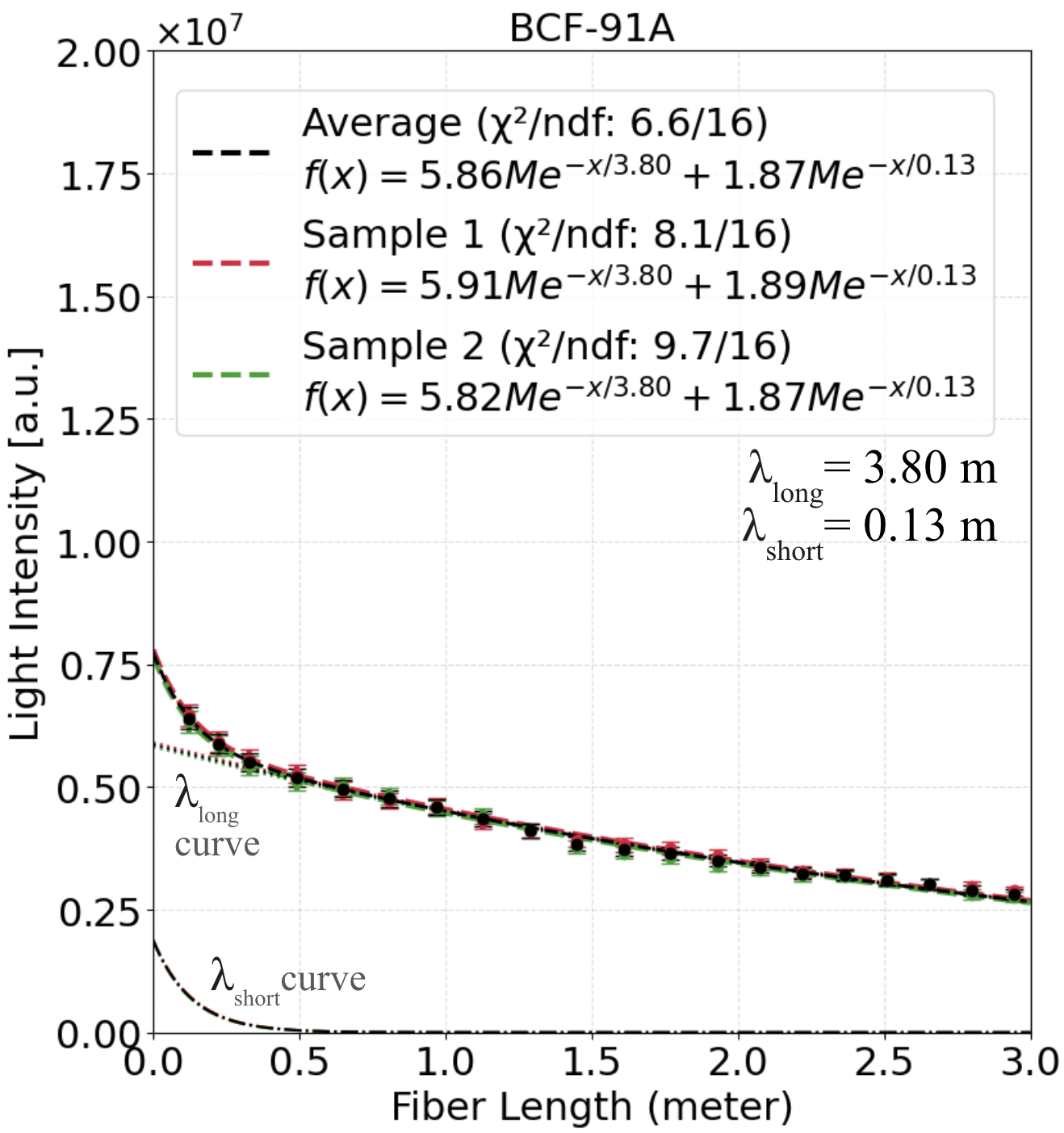}
    \includegraphics[width=0.495\textwidth]
    {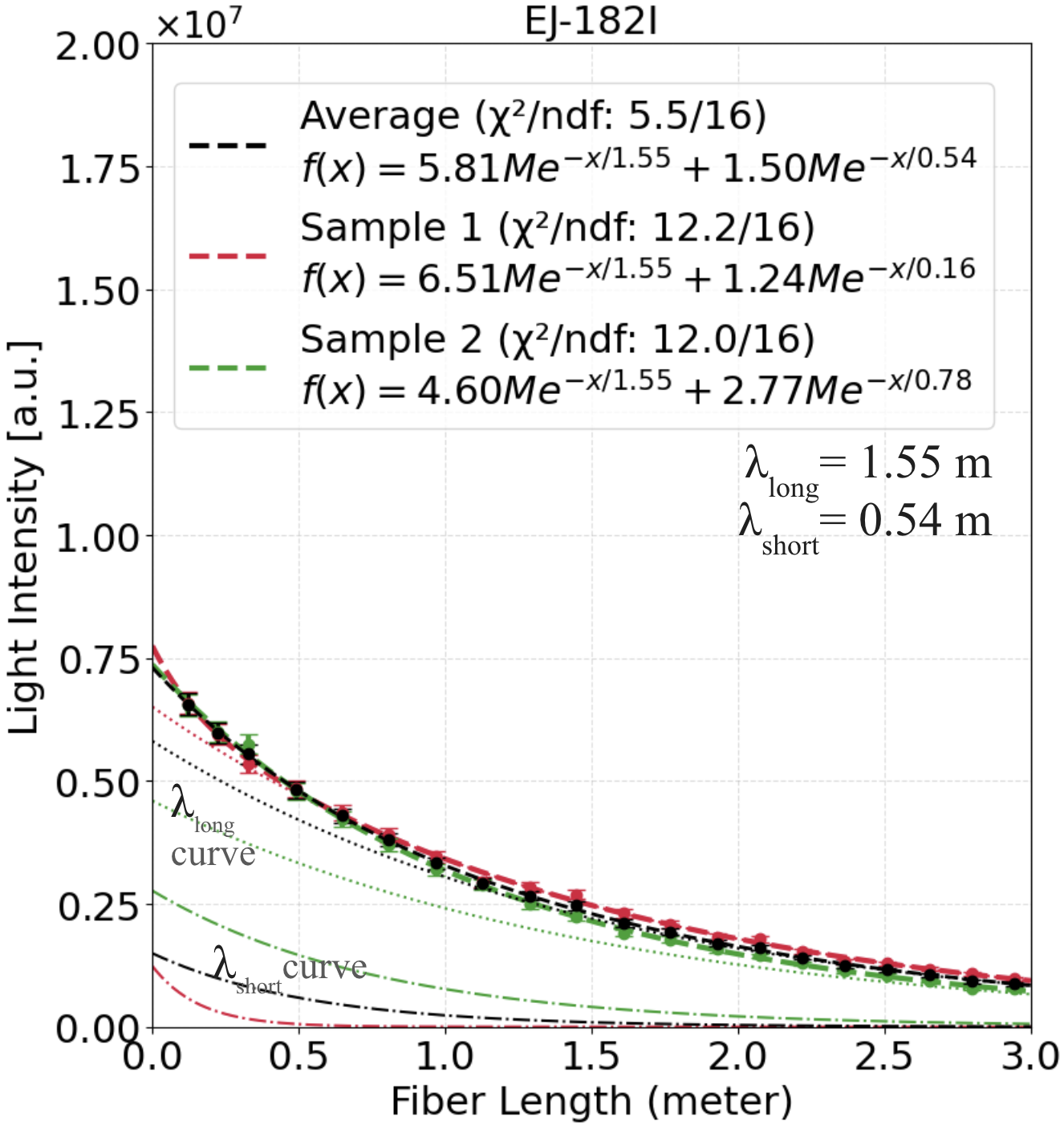}

    \vspace{1.0em} 
    
    %2nd line 
    \includegraphics[width=0.495\textwidth]
    {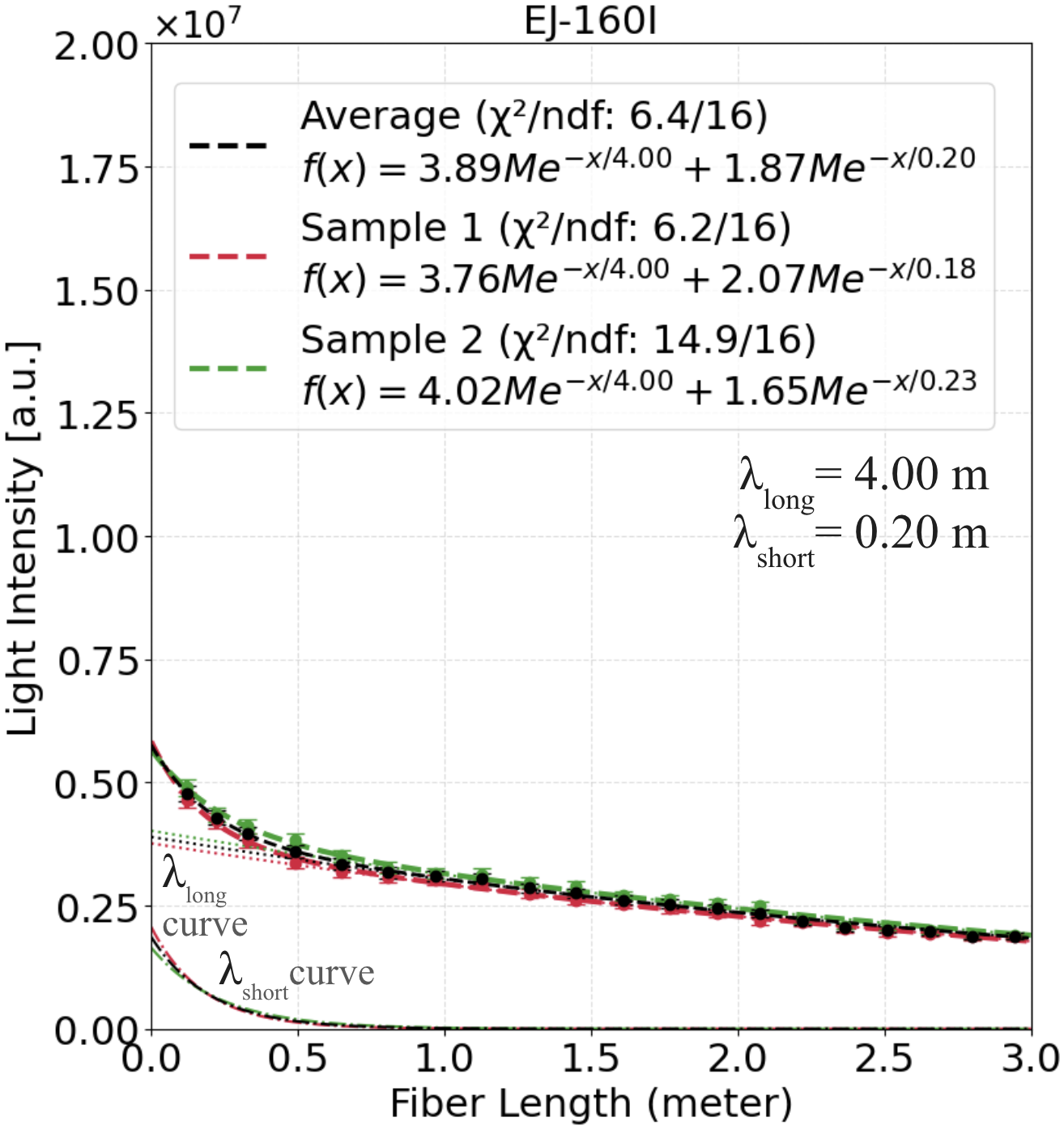}
    \includegraphics[width=0.495\textwidth]
    {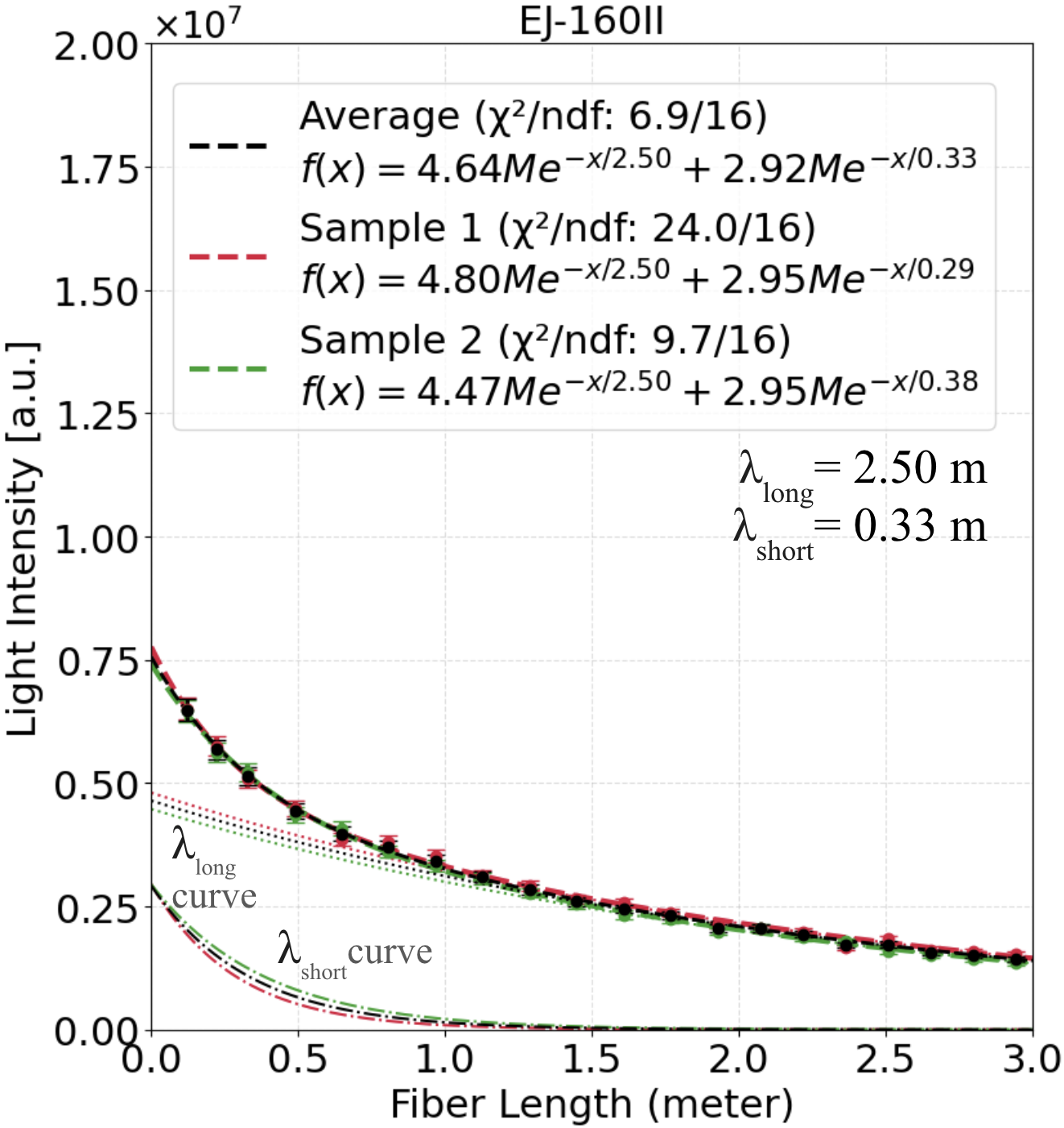}

    \caption{Light intensity as a function of the light propagation distance for various fibers: BCF-91A (top left), EJ-182I (top right), EJ-160I (bottom left), and EJ-160II (bottom right). Each plot shows results for two fiber samples of the corresponding type. The data points are fitted with double-exponential functions according to formula~\ref{formula}. The long and short components are also indicated by separate dashed lines.
    }

    \label{fig: integrated attenuation length2}
\end{figure}

The light intensities shown in Figures~\ref{fig: integrated attenuation length2} were obtained by integrating the measured emission spectra in the range 450–700 nm; 
see Figure~\ref{fig: emission spectrum of 4 green WLS fibers}. 
In this work, “light output” refers to the overall amount of collected light, while “light intensity” is used when this quantity is defined from the measured emission spectrum (for example, by integrating the spectrum over wavelength).
Each data point corresponds to the average of 10 consecutive intrinsic measurements from the spectrophotometer, each integrated over 10 ms of light collection, resulting in statistically negligible uncertainty.
%
%The systematic uncertainty was estimated to be 3.5\,\% per data point, based on repeated independent measurements in which both the fiber and the LED were repositioned between trials.
%
The systematic uncertainty was estimated to be 3.5\,\% per data point, based on the variation of the integrated emission in repeated independent measurements in which the fiber and the LED were repositioned between trials, three times for the fiber and twenty times for the LED.
To ensure measurement reliability, the LED output was independently verified for each round of measurement using a separate spectrophotometer, and the resulting fiber light output was linearly normalized to account for any deviation in LED intensity.

The light attenuation is modeled by a double-exponential function of the form:
\begin{equation}
I = I_{\text{long}}\,e^{-x/\lambda_{\text{long}}} + I_{\text{short}}\,e^{-x/\lambda_{\text{short}}},
\label{formula}
\end{equation}
where $I$ is the light intensity recorded by the spectrophotometer, 
and $x$ is the distance between the LED source and the spectrophotometer.
$I_{\text{long}}$, $I_{\text{short}}$, $\lambda_{\text{long}}$, and $\lambda_{\text{short}}$ denote long and short components of the light intensities and attenuation lengths.
It is common in the literature and various applications that only $\lambda_{\text{long}}$ is used as ``the fiber attenuation length''. However, at shorter distances a double-exponent function provides a much better description of the light intensities.~\cite{MINOS-Michael:2008bc,Avvakumov:2005ww}.

\begin{figure}[h!]
    \centering
    \includegraphics[width=0.55\textwidth]{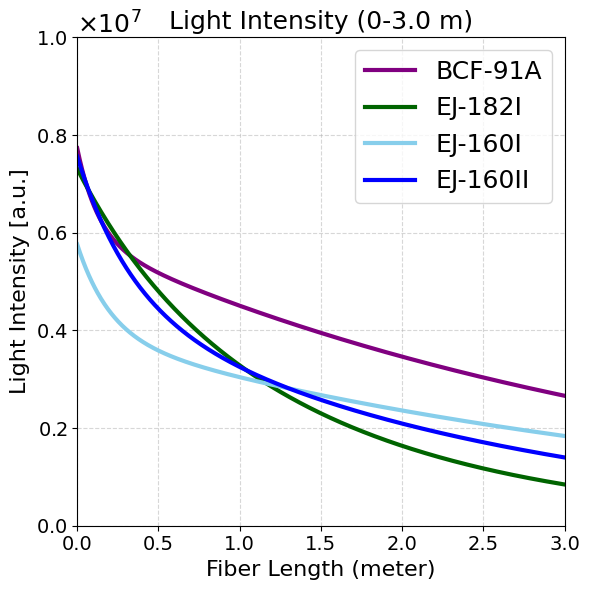}
    \caption{Comparison of fitted light intensities of the four fibers.}
    \label{fig: Light-Yield Comparision1}
\end{figure}

The long attenuation lengths fitted ($\lambda_{\text{long}}$) are 3.80\,m for BCF-91A, 1.55\,m for EJ-182I, 4.00\,m for EJ-160I, and 2.50\,m for EJ-160II. The value of $\lambda_{\text{long}} = 3.80$\,m for BCF-91A is consistent with the manufacturer's specification and with previous internal measurements by the LEGEND collaboration~\cite{Krause, LEGEND-LAr-EPJ2021}. 
Figure~\ref{fig: Light-Yield Comparision1} shows the comparison of the fitted light intensities of the four fibers as shown in Figures~\ref{fig: integrated attenuation length2}. The summary of these results and the integrated light intensity over 1.40\,m and 3.00\,m are displayed in Table~\ref{tab: light output table}. Here, 1.40\,m corresponds to the fiber length foreseen for the LEGEND-1000 design~\cite{LEGEND:2021bnm}, while 3.00\,m matches the actual sample length used in our measurements. 
%As expected, the EJ-160 variants show that a higher fluor concentration results in a higher light output but a shorter $\lambda_{\text{long}}$ due to self-absorption. 
%==================================================================================

%==================================================================================

\clearpage

% ---------------- table ----------------

\begin{table}[h!]
    \centering
    \begin{tabular}{|>{\centering\arraybackslash}m{2.0cm}|c|c|>{\centering\arraybackslash}m{3cm}|>{\centering\arraybackslash}m{3cm}|}
        \hline
        Fiber & $\lambda_{\text{long}}$ (m) & $\lambda_{\text{short}}$ (m) & \makecell{Integrated \\ light intensity \\ (0\,m -- 1.40\,m)} & \makecell{Integrated \\ light intensity \\ (0\,m -- 3.00\,m)} \\
        \hline
        \raggedright BCF-91A      & 3.80 ± 0.11 & 0.13 ± 0.04
        & 1.00 & 1.00 \\
        \raggedright EJ-182I       & 1.55 ± 0.11 & 0.54 ± 0.19 & 0.86 & 0.69 \\
        \raggedright EJ-160I       & 4.00 ± 0.15 & 0.20 ± 0.03 & 0.70 & 0.69 \\
        \raggedright EJ-160II      & 2.50 ± 0.12 & 0.33 ± 0.04 & 0.83 & 0.73 \\
        \hline
    \end{tabular}
    \caption{Long and short attenuation lengths, and relative integrated light intensities over 0\,m -- 1.40\,m (LEGEND-1000 design fiber length~\cite{LEGEND:2021bnm}) and 0\,m -- 3.00\,m (sample length used in this study).}
    \label{tab: light output table}
\end{table}

%==================================================================================

% \clearpage

%==================================================================================
%==================================================================================
\subsection{Spectral attenuation lengths}

In Figure~\ref{fig: emission spectrum of 4 green WLS fibers}, all fibers show a gradual shift of emission toward longer wavelengths with increasing \textit{light propagation distance}. This is accompanied by a rapid decrease in light intensity at short wavelengths and a comparatively slower decay at longer wavelengths, consistent with previously reported trends for the Kuraray Y-11 WLS fiber~\cite{Pahlka2019}. 
This is a clear observation of the effect that the attenuation length is wavelength-dependent, despite the common practice of quoting a single constant value.

\begin{figure}[h!]
    \centering
    \includegraphics[width=.80\textwidth]{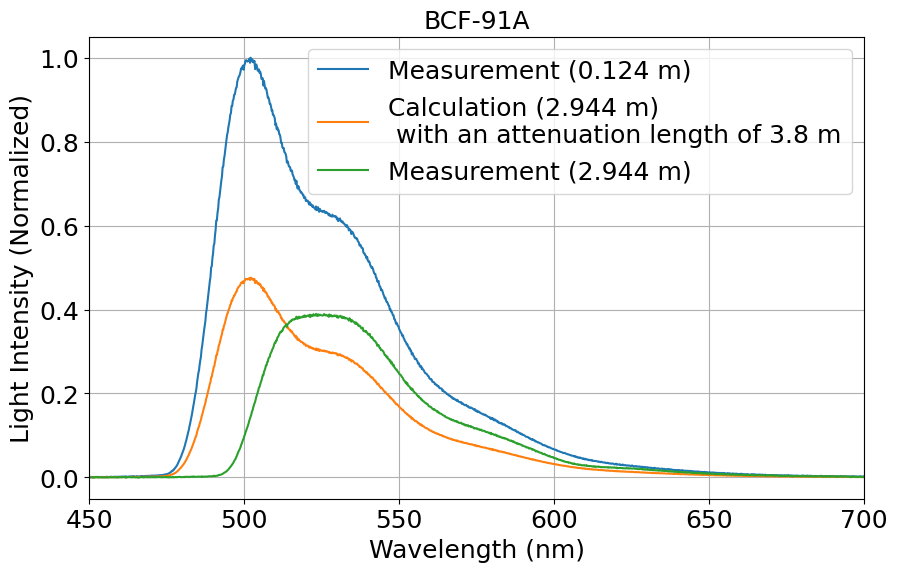}
    \caption{\raggedright
    Emission spectra for BCF-91A measured at 0.124\,m (blue); expected at 2.944\,m assuming 
    a constant attenuation length of 3.8\,m (orange); and measured at 2.944\,m (green).
    }
    \label{fig: schematic diagram of light transport}
\end{figure}

Figure~\ref{fig: schematic diagram of light transport} illustrates this effect for the BCF-91A fiber. 
The blue curve corresponds to the spectrum measured at a light-propagation distance of 0.124\,m, and the green curve represents the spectrum measured at 2.944\,m. These two spectra correspond to those shown for BCF-91A in 
Figure~\ref{fig: emission spectrum of 4 green WLS fibers}. 
The orange curve shows the expected spectrum at 2.944\,m assuming a constant attenuation length of 3.8\,m applied to the 0.124\,m data. The discrepancy between the orange and green curves clearly demonstrates the presence of wavelength-dependent attenuation.

%==================================================================================
%==================================================================================
%==================================================================================
%==================================================================================

\begin{figure}[h!]
\centering
\includegraphics[width=.495\textwidth, height=.495\textwidth]
{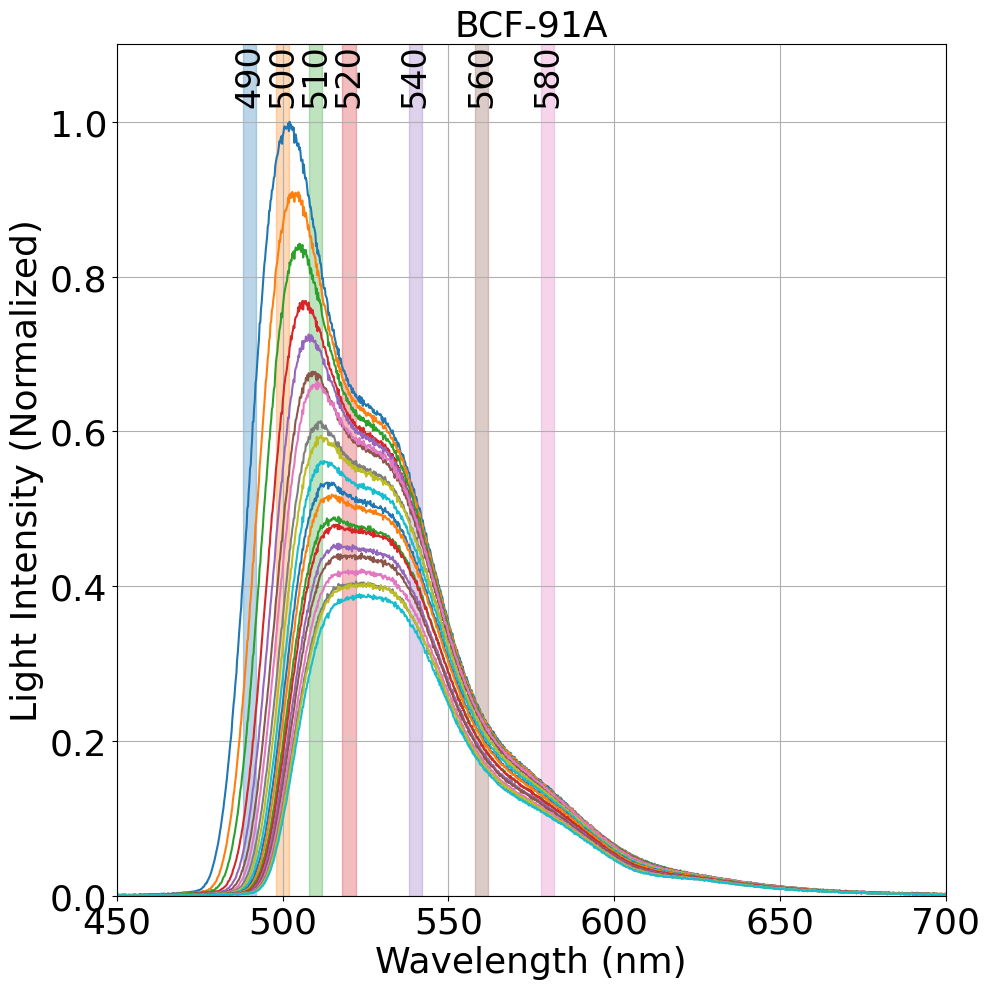}
\includegraphics[width=.495\textwidth, height=.495\textwidth]
{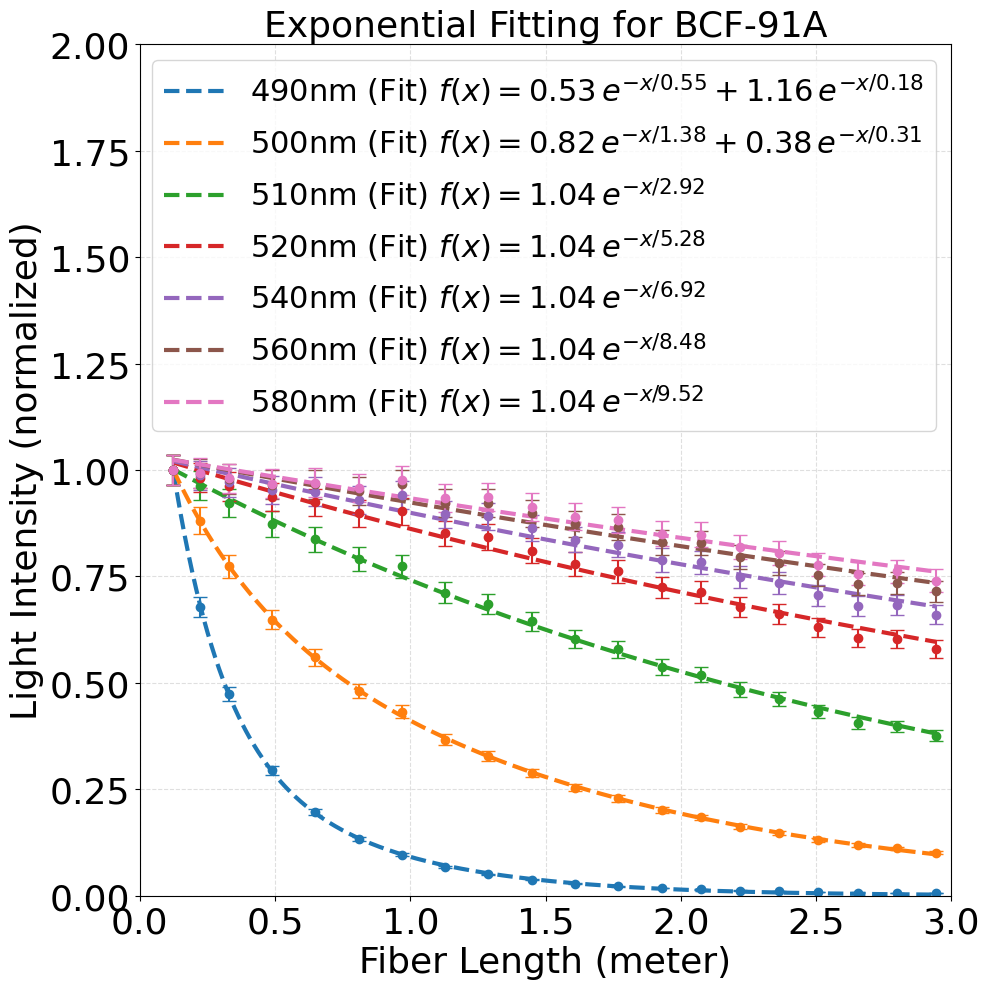}
\caption{\raggedright 
(Left) Emission spectra for BCF-91A with seven selected wavelength regions indicated, as in Figure~\ref{fig: emission spectrum of 4 green WLS fibers}. 
(Right) Exponential fits to each region, used to extract
$\Lambda_{\text{long}}(\lambda)$ and $\Lambda_{\text{short}}(\lambda)$ values.
For 490 and 500\,nm we use a sum of short and long exponentials, whereas for 510--580\,nm a single exponential is sufficient. Accordingly, only the long component is quoted for 510--580\,nm in Table~\ref{tab: attenuation length with selected wavelength table}, and no short component is listed for these wavelengths.
}

\label{fig: attenuation length with selected wavelength} 
\end{figure}

Quantifying wavelength-dependent attenuation lengths $\Lambda_{\text{long}}(\lambda)$ and $\Lambda_{\text{short}}(\lambda)$, we fit the measured light intensity to the formula:
\begin{equation}
I(\lambda) = I_{\text{long}}(\lambda)\,e^{-x/\Lambda_{\text{long}}(\lambda)} + I_{\text{short}}(\lambda)\,e^{-x/\Lambda_{\text{short}}(\lambda)}.
\label{eq: attenuation2}
\end{equation}
This model was first applied to seven discrete wavelength bands, shown in Figure~\ref{fig: attenuation length with selected wavelength}, with results summarized in Table~\ref{tab: attenuation length with selected wavelength table}. A $\pm 2.0$\,nm band width was chosen to ensure sufficient statistics in each fit. To assess whether both components are statistically required, we compared single- and double-exponential fits using a $\chi^{2}$ difference test ($\Delta\mathrm{dof}=2$). For 490 and 500\,nm, the two-component model is clearly favored ($\Delta\chi^{2} \gg 6$, corresponding to $p \ll 0.05$), whereas for 510--580\,nm the improvement in $\chi^{2}$ is negligible ($\Delta\chi^{2} \simeq 0$, $p \simeq 1$). We therefore regard the fitted short-component amplitudes above 510\,nm as not statistically significant and quote only the long component in Table~\ref{tab: attenuation length with selected wavelength table}.

\begin{figure}[h!]
\centering
\includegraphics[width=.495\textwidth]{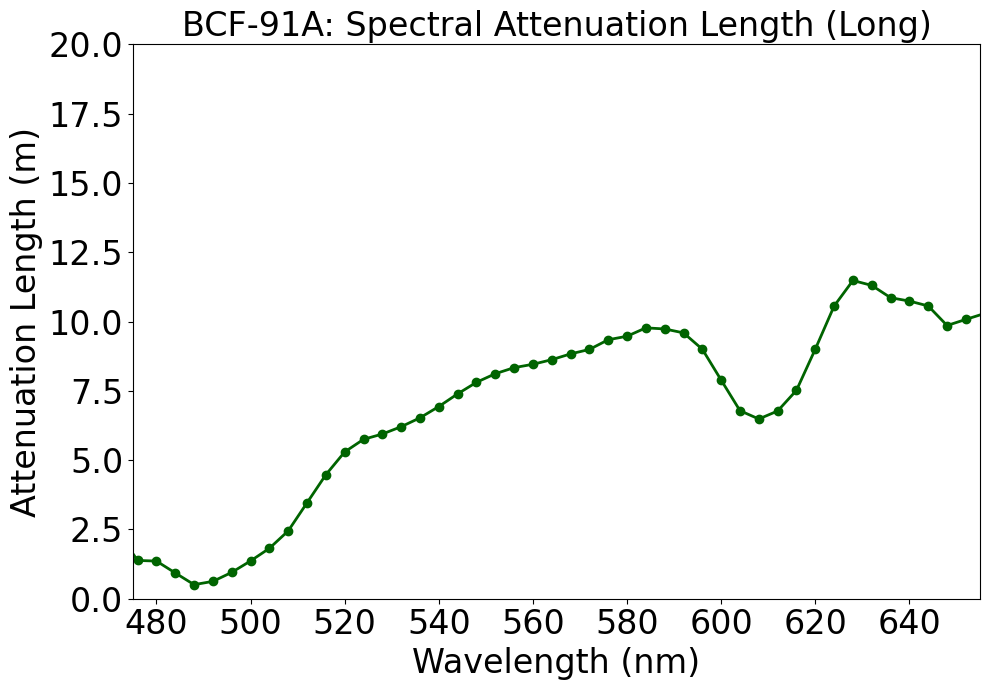}
\includegraphics[width=.495\textwidth]{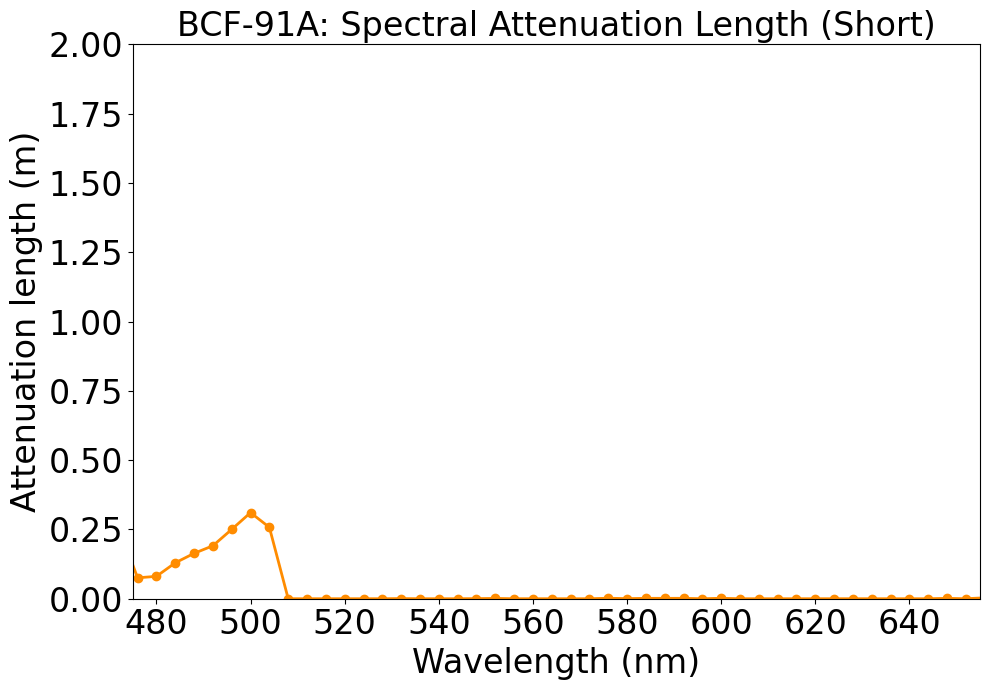}
\includegraphics[width=.495\textwidth]{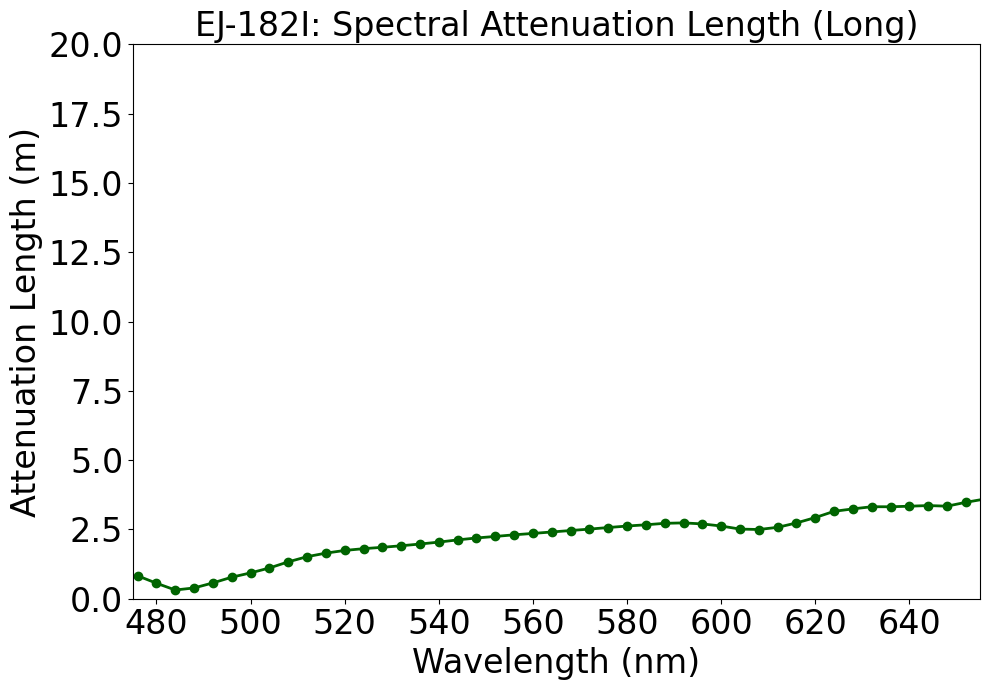}
\includegraphics[width=.495\textwidth]{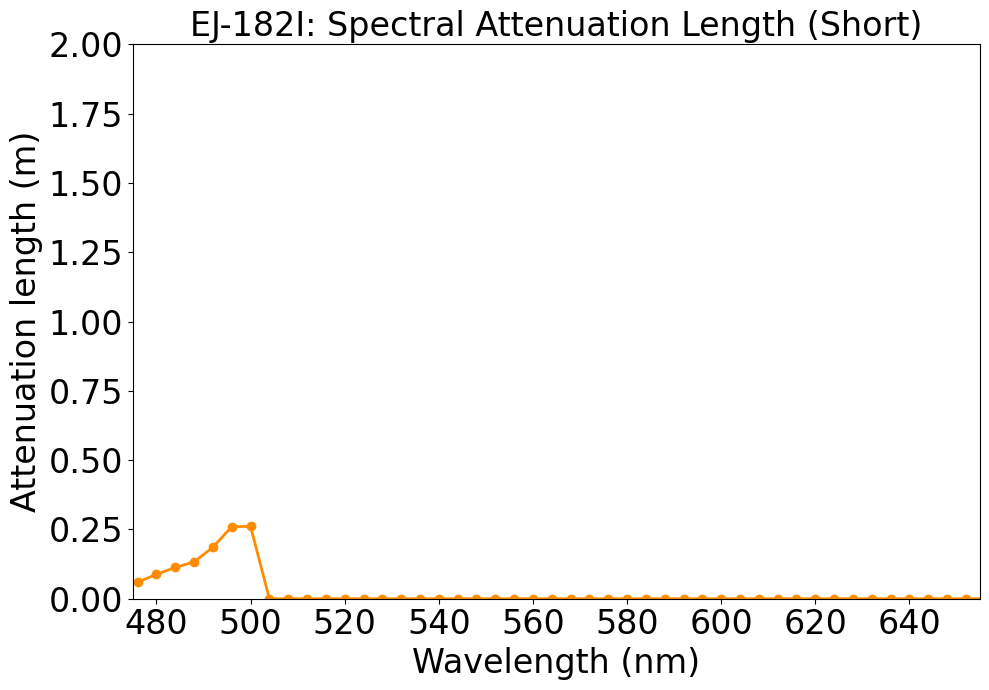}
\includegraphics[width=.495\textwidth]{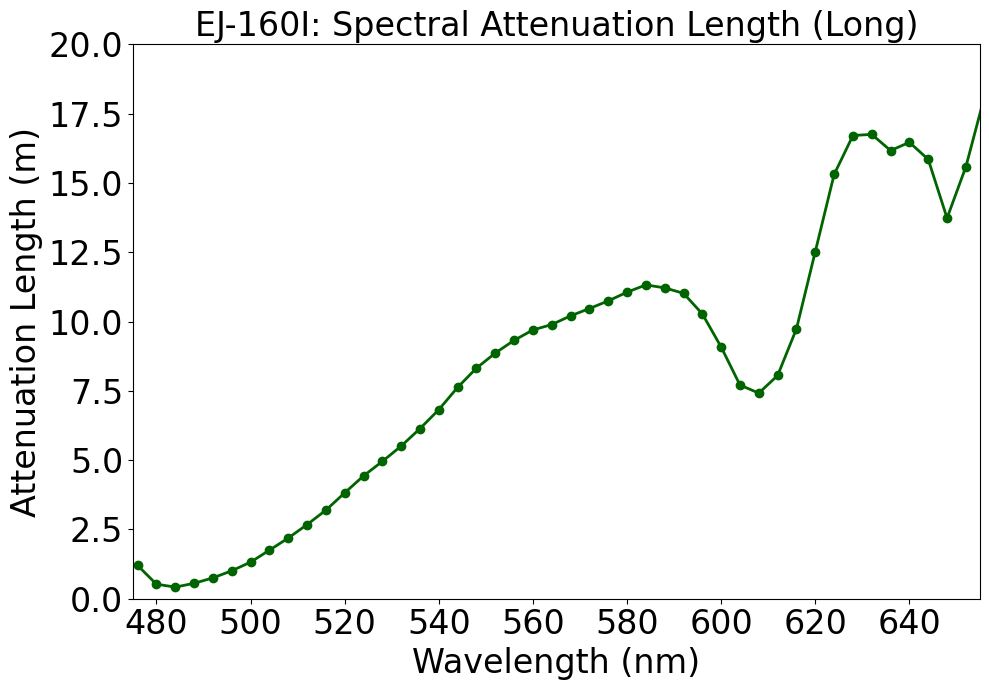}
\includegraphics[width=.495\textwidth]{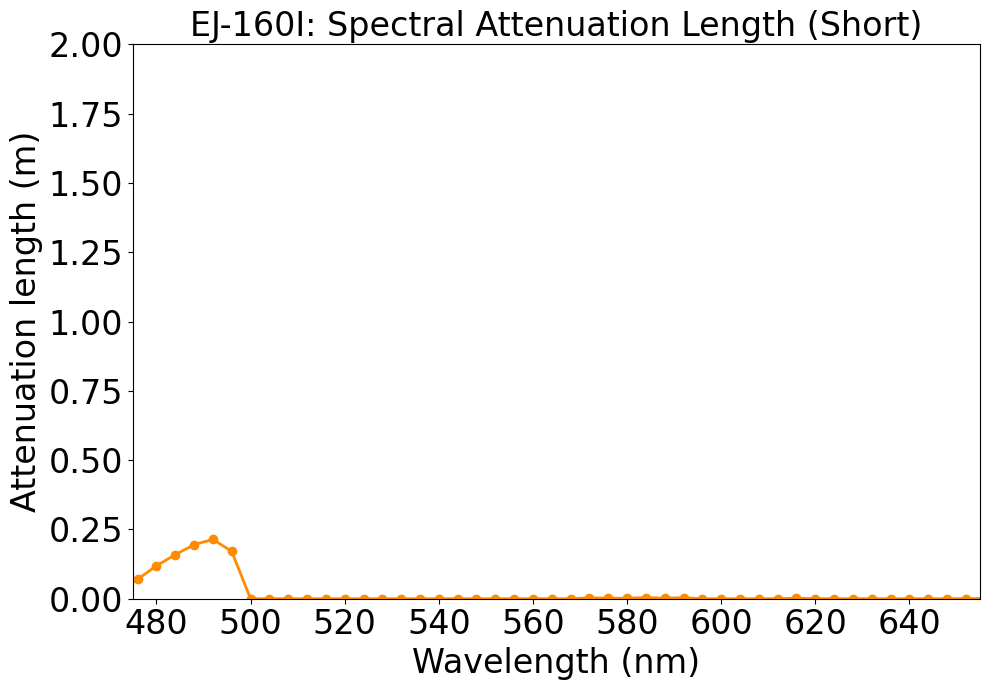}
\includegraphics[width=.495\textwidth]{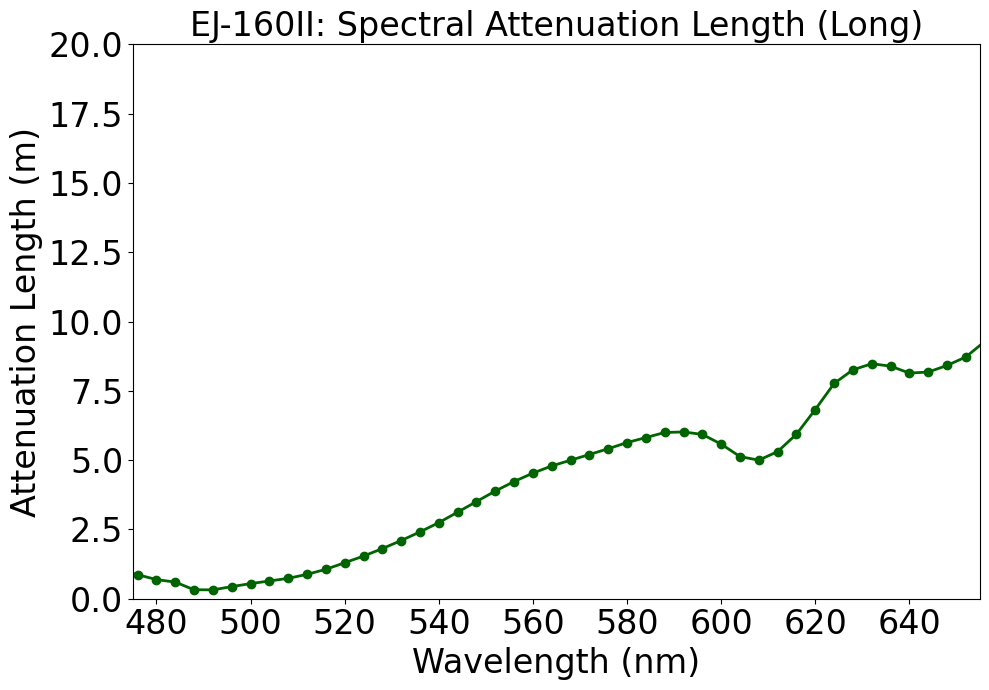}
\includegraphics[width=.495\textwidth]{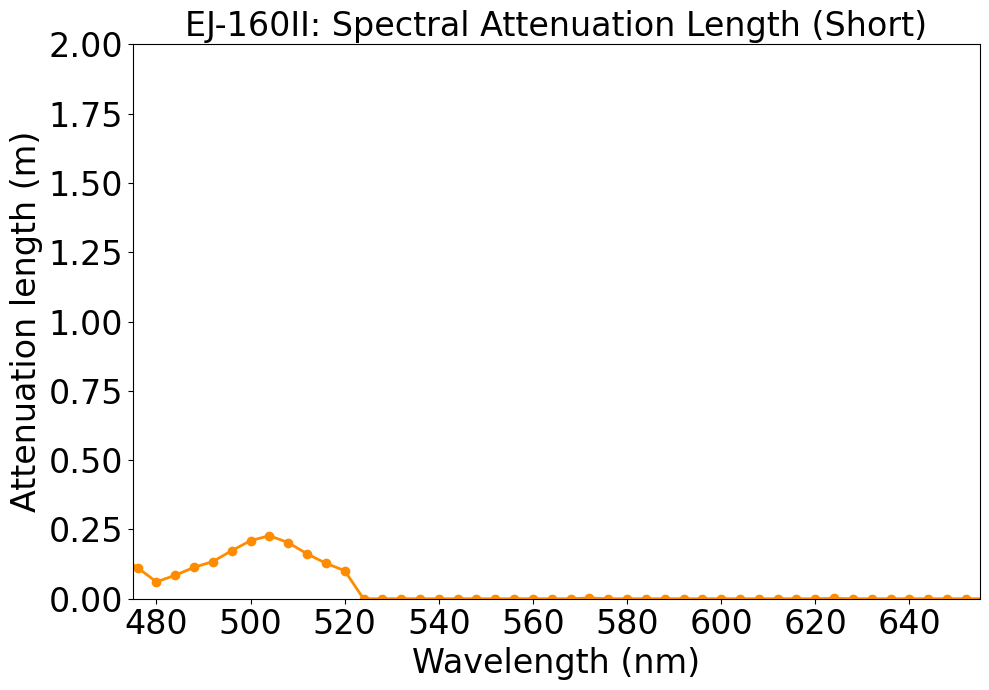}

\caption{Spectral attenuation lengths for BCF-91A, EJ-182I, EJ-160I, and EJ-160II. The left column shows $\Lambda_{\text{long}}(\lambda)$ and the right column shows $\Lambda_{\text{short}}(\lambda)$.}

\label{fig: spectral attenuation length}
\end{figure}

Extending the analysis over the range 475 to 655\,nm (beyond which the signal falls below 1\% of the peak) gives the spectral attenuation curves shown in Figure~\ref{fig: spectral attenuation length}. As expected, $\Lambda_{\text{long}}(\lambda)$ increases with wavelength and exhibits localized dips near 490, 610 and 650\,nm, consistent with previously reported trends for Kuraray Y-11 WLS fibers~\cite{Mu2e-JINST}.
For $\Lambda_{\text{short}}(\lambda)$ (typically $<$1.0\,m), to avoid bias from poorly constrained fits resulting from uncertainties in low-intensity regions, the values are reported only when the fraction of short components satisfies the condition $I_{\text{short}}/(I_{\text{long}} + I_{\text{short}}) > 0.03$. This threshold improves the stability of the fit and confirms that short-wavelength attenuation becomes negligible above the main emission peak ($>$500--520\,nm).

%==================================================================================
%==================================================================================

%==================================================================================
%==================================================================================
\begin{table}[h!]
\centering
\begin{tabular}{|c|c|c|}
\hline
Wavelength (nm) & $ \Lambda_{\text{long}}(\lambda) $ (m) & $ \Lambda_{\text{short}}(\lambda) $ (m) \\
\hline
490 & 0.55 & 0.18 \\
500 & 1.38 & 0.31 \\
510 & 2.92 & - \\
520 & 5.28 & - \\
540 & 6.92 & - \\
560 & 8.48 & - \\
580 & 9.52 & - \\
\hline
\end{tabular}
\caption{Fitted attenuation lengths $\Lambda_{\text{long}}(\lambda)$ and $\Lambda_{\text{short}}(\lambda)$ for BCF-91A in seven selected wavelength regions. A dash (-) indicates that only the long component is quoted because the short component is not statistically required in that wavelength region.}
\label{tab: attenuation length with selected wavelength table}
\end{table}
%==================================================================================
%==================================================================================

\clearpage

%==================================================================================
%==================================================================================
\subsection{Measurements in water}
\label{sec:measurement_in_water}

In practical applications, fibers may be surrounded by other media than air such as a solidified glue (e.g., in the MINOS experiment~\cite{MINOS-Michael:2008bc}),  liquid scintillator (e.g., in the NOvA experiment~\cite{Ayres:2004js}), liquid argon (e.g., in the LEGEND experiment~\cite{LEGEND:2025insdet, LEGEND:2021bnm}), or other environments.
This prompted us to conduct measurements with fibers immersed in water with an index of refraction much closer to some of the above examples.

Since the propagation of light in fibers is governed by the total internal reflection at the core–cladding and the cladding–environment interfaces, then the fraction of light that escapes the fibers increases as the difference in magnitudes of the indices of refraction decreases. This occurs when the incident angle exceeds the \textit{critical angle} $\theta_c$:
\[
\theta_c = \arcsin\left(\frac{n_2}{n_1}\right), \quad \text{for } n_1 > n_2,
\]
where \(n_1\) and \(n_2\) are the refractive indices of the adjacent media.
For reference, the refractive index of air is \(n \approx 1.0003\), liquid argon at 86\,K is \(n \approx 1.23\)~\cite{Rindex_LAr,Babicz2020}, and water is \(n \approx 1.335\)~\cite{Rindex_Water}. Using PMMA as the cladding material with $n\approx 1.49$, the corresponding critical angles are approximately \(42.3^\circ\) (air), \(55.8^\circ\) (liquid argon) and \(63.8^\circ\) (water), implying increased light leakage as the surrounding refractive index increases. 
For consistency, all values are quoted at the wavelength of 500\,nm, near the emission peak of the WLS fibers.

\begin{figure}[h!]
\centering
\includegraphics[width=0.7\textwidth]{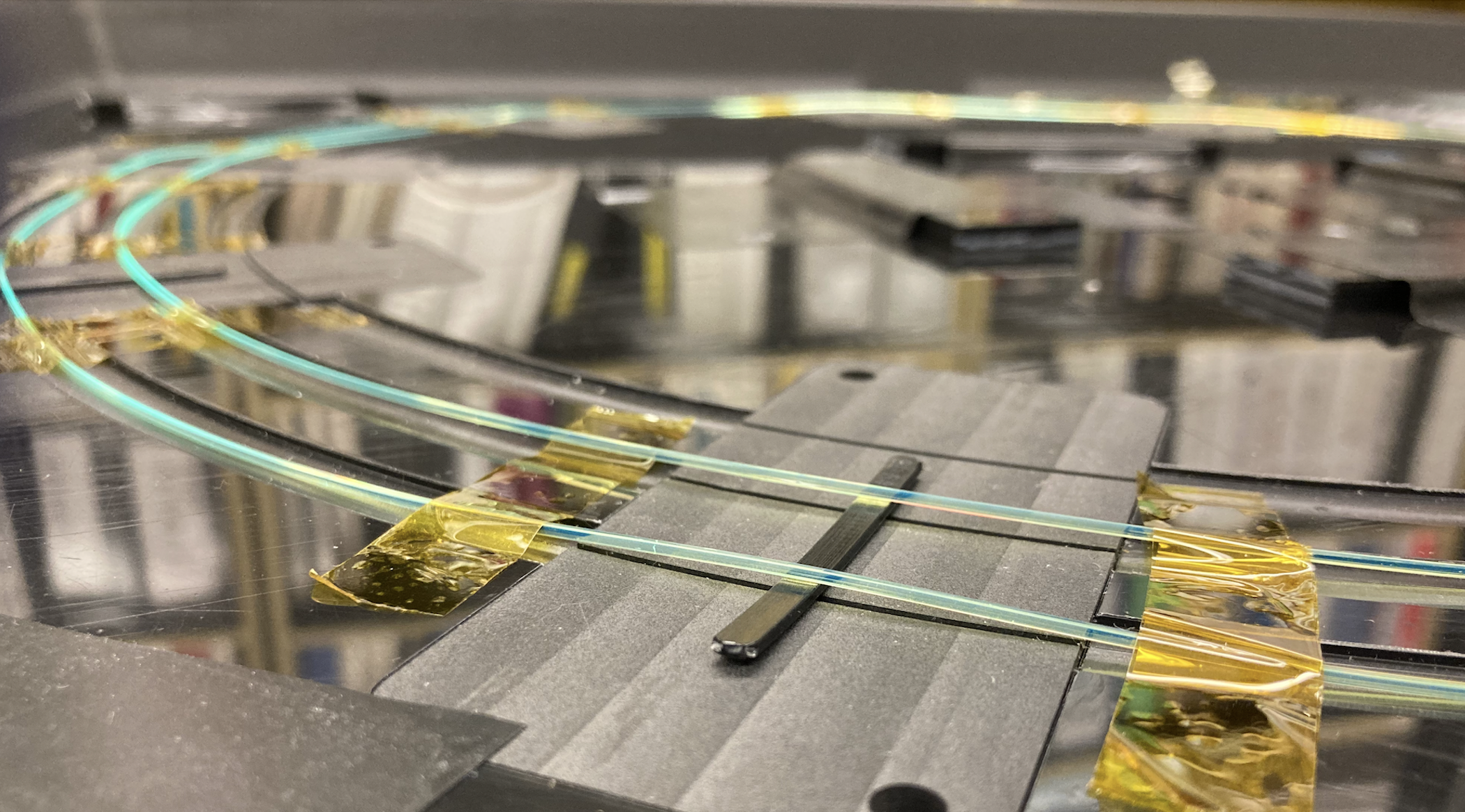}

\vspace{0.5cm}
\includegraphics[width=0.71\textwidth]{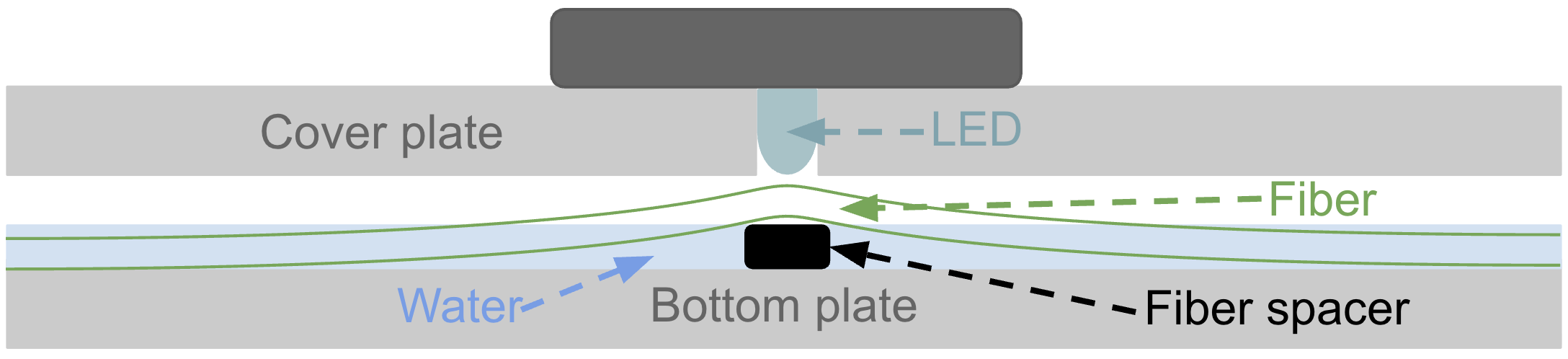}

\caption{\raggedright 
Test configuration for measurements in water (top) and its schematic diagram (bottom).
The fiber is partially elevated 
in order to expose a segment of the fiber that can be illuminated by an LED without submersing the LED.
While this arrangement introduces partial non-immersion boundaries and minor geometric deviations, it allows for a controlled study of refractive-index effects at the air/water interface.}
\label{fig:optical bench in water}
\end{figure}

\begin{figure}[h!]
\centering
\begin{tabular}{cc}
\includegraphics[width=.49\textwidth]{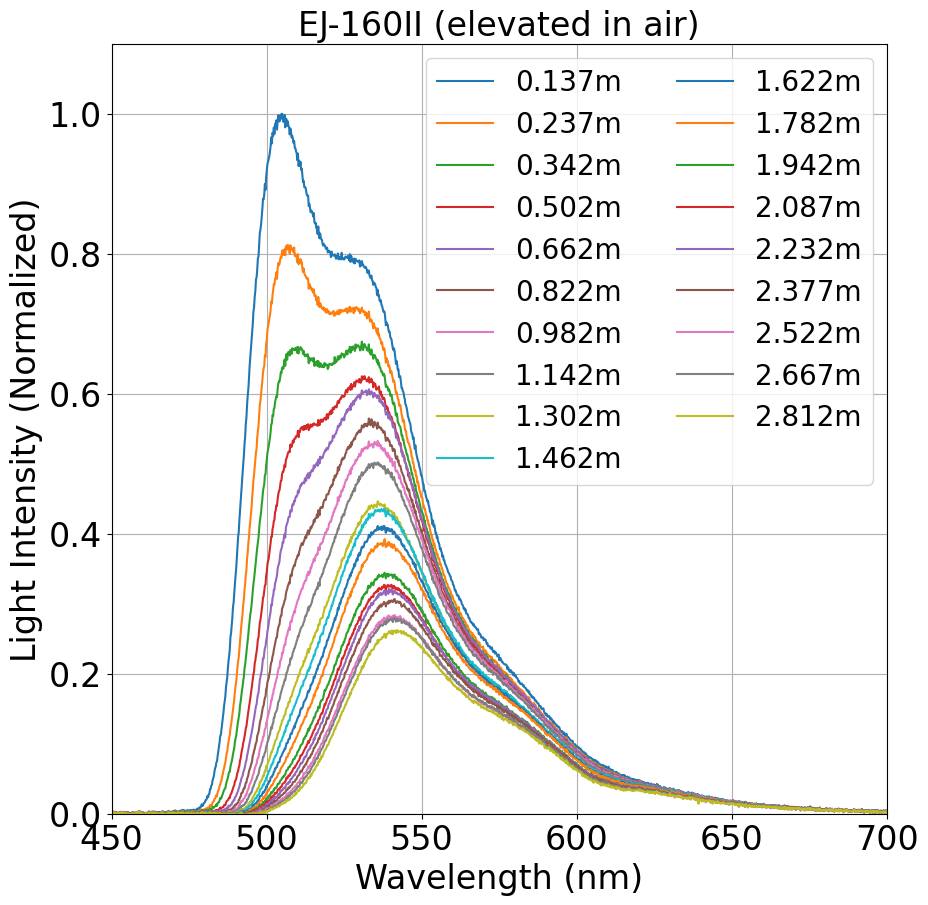} &
\includegraphics[width=.49\textwidth]{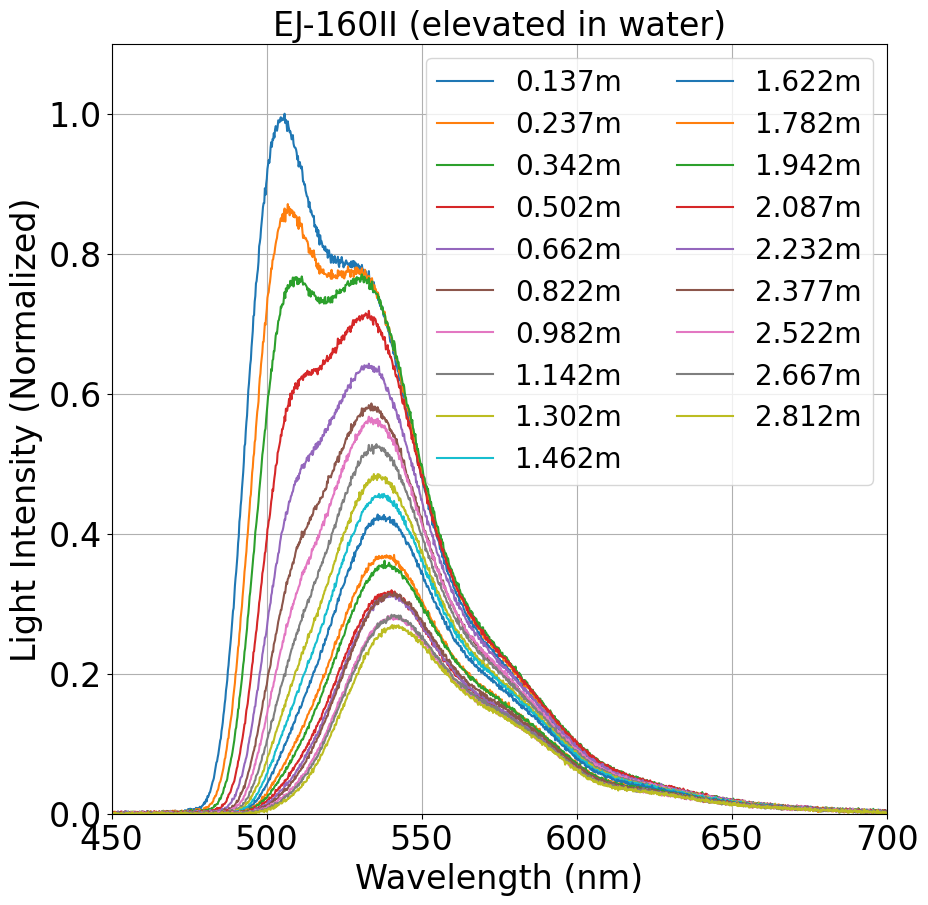}
\end{tabular}
\captionsetup{justification=centering}
\caption{Normalized emission spectra of EJ-160II: \\ elevated in air (left) and elevated in water (right).}

\label{fig: emission spectrum of EJ-160II air and water}
\end{figure}

Figure~\ref{fig:optical bench in water} shows the plate with a groove for testing fibers in water, as well as its schematic diagram.
In this setup, the fiber was elevated by about 0.05\,inch near the holes, while the rest of the fiber was fully immersed in water. Measurements were taken only for the EJ-160II fiber. To ensure a fair comparison, air measurements were repeated with slightly modified (``elevated'') fiber positions near the LED holes and then with the rest of the fiber immersed in water. 
The results of these measurements are shown in Figure~\ref{fig: emission spectrum of EJ-160II air and water}.

%==================================================================================
%==================================================================================

\begin{figure}[h!]
    \centering    
    \includegraphics[height=0.49\textwidth]{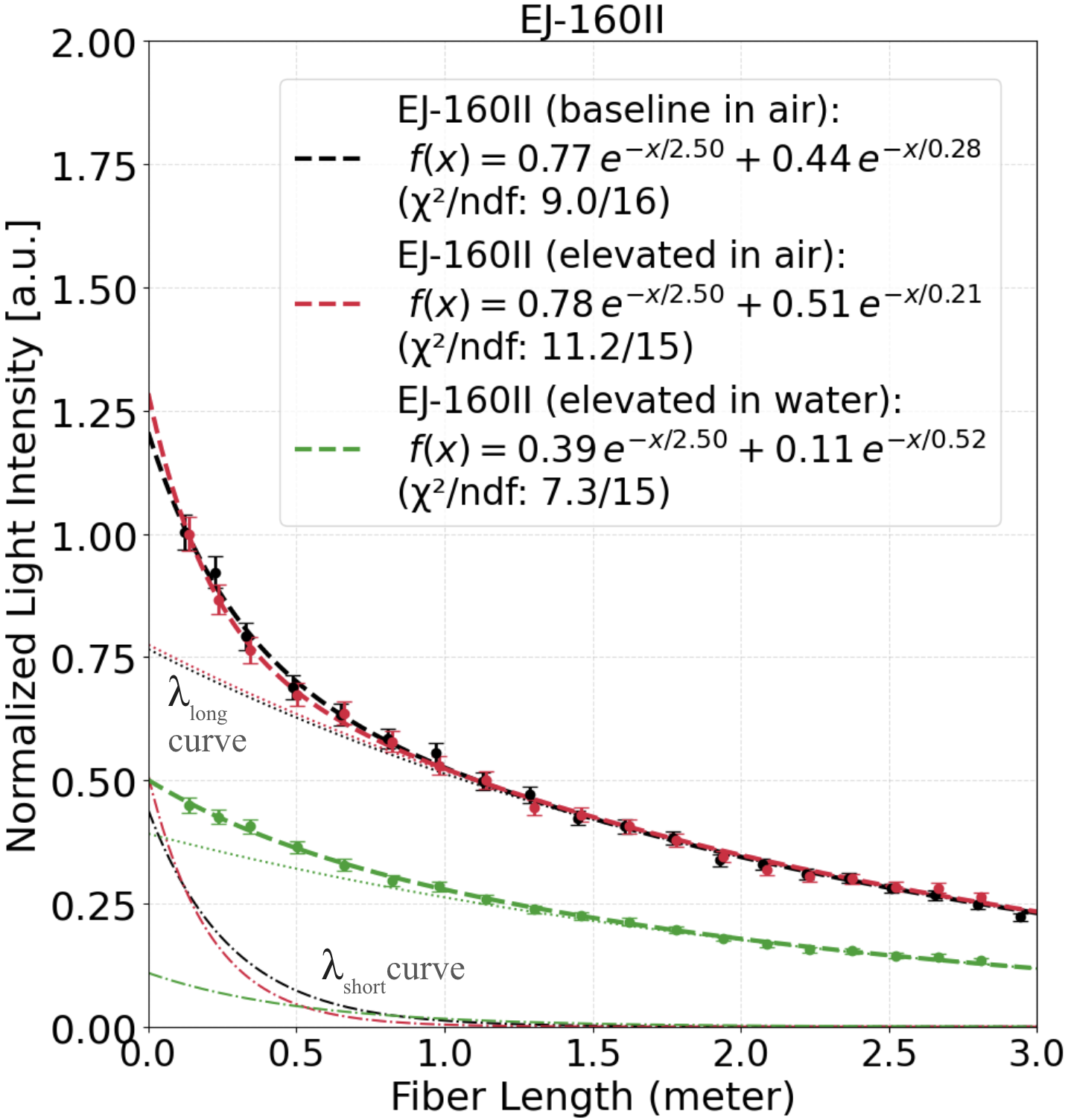}
    \caption{\raggedright Light intensity decay of EJ-160II under three conditions: 
    (1) baseline in air, (2) elevated in air, and (3) elevated in water.
    The data points are fitted with double-exponential functions according to formula~\ref{formula}. The long and short components are also indicated by separate dashed lines.
    }
    \label{fig: integrated attenuation length in water}
    \end{figure}

%==================================================================================
%==================================================================================
Figure~\ref{fig: emission spectrum of EJ-160II air and water} compares the emission spectra for EJ-160II in air and in water in the modified setup. The data in water exhibit slower light intensity decay at short distances compared to the data in air, which is also reflected in Figure~\ref{fig: integrated attenuation length in water}. Figure~\ref{fig: integrated attenuation length in water} shows the light intensity in the EJ-160II fiber under three conditions: (1) baseline in air, (2) elevated in air, and (3) elevated in water. For effective comparison, the light intensity was normalized such that the highest value of condition (2), elevated in air, was set to unity. The fits for conditions (1) and (2) are nearly identical, indicating that the modification of the setup introduced a negligible optical effect. 
Submerging the fiber in water reduces the integrated light intensity in the 0–3.0\,m region to approximately 49\,\% of that in the air. In addition, the contribution of the short attenuation component ($I_{\text{short}}$) is strongly suppressed relative to the long component ($I_{\text{long}}$).

%==================================================================================
%==================================================================================

%==================================================================================
%==================================================================================

%==================================================================================
%==================================================================================
\clearpage

%==================================================================================
%==================================================================================

\section{Conclusion}

This study presents an optical characterization of the new fibers EJ-182I, EJ-160I, and EJ-160II from Eljen Technology and their comparison with a BCF-91A fiber from Saint Gobain (now Luxium Solutions). The fitted long attenuation lengths ($\lambda_{\text{long}}$) were found to be 3.80~m for BCF-91A, 1.55~m for EJ-182I, 4.00~m for EJ-160I, and 2.50~m for EJ-160II. Among the EJ-160 variants, EJ-160I shows longer attenuation length with lower light intensity, and EJ-160II shows shorter attenuation length with higher light intensity.

Spectral attenuation analysis revealed a strong wavelength dependence of attenuation. The long attenuation length $\Lambda_{\text{long}}(\lambda)$ increased with wavelength and exhibited localized dips near 490, 610 and 650\,nm. The short attenuation length $\Lambda_{\text{short}}(\lambda)$ became negligible above 500--520\,nm.

Tests with a EJ-160II fiber immersed in water show a reduction in the overall light intensity and suppression of the short attenuation component ($I_{\text{short}}$), consistent with reduced refractive index contrast between the fiber cladding and the surrounding medium.

%can I add this one? beacuse we have other WLS fibers we have not shown here. 0.7mm circular, 1mm circular... which we deleted in the appendix.
%This publication presents partial results of the ongoing program to develop better WLS and Sci-WLS fibers and improve their radio purity.
%
% Future work includes extending these measurements to additional fiber variants that are currently in development at Eljen Technology. 
% %
% We are also advancing a comprehensive simulation package for light output and propagation in WLS fibers and will report on the modeling results in future publications.

% Future work includes extending these measurements to additional fiber variants that are currently in development at Eljen Technology.
% We are also advancing a comprehensive simulation package for light output and propagation in WLS fibers, a similar approach having been previously validated for Kuraray Y-11 fibers~\cite{Pahlka2019}, and will report on the modeling results in future publications.

Future work includes extending these measurements to additional fiber variants that are currently in development at Eljen Technology, as well as to fibers with different cross-sections.
%
%We are also advancing a comprehensive simulation package for light output and propagation in WLS fibers, a similar approach having been previously validated for Kuraray Y-11 fibers~\cite{Pahlka2019}, and will report on the modeling results in future publications.
%
We are also advancing a comprehensive ray-tracing simulation of light output and propagation in WLS and Sci-WLS fibers, in which optical photons are tracked as rays using a photon-transport Monte Carlo approach previously validated for Kuraray Y-11 fibers~\cite{Pahlka2019}. This simulation forms part of a broader modeling program that builds on and extends the measurements reported in this work, and we plan to describe the simulation framework and results in a separate publication.

%==================================================================================
%==================================================================================
\acknowledgments{
We thank Prof.~S.~Schönert, Dr.~P.~Krause, and the group at the Technical University of Munich for providing BCF-91A fiber samples and for insightful discussions. 
This work was supported in part by the University of Texas at Austin and the U.S. National Science Foundation under grant PHY-2312278.
}
%==================================================================================
%==================================================================================

% \clearpage

%==================================================================================
%==================================================================================
% \nocite{*}  % 모든 references.bib 항목 포함
% \bibliographystyle{JHEP}
% \bibliography{images_and_references/gen/references}
% \bibliographystyle{JHEP}  % 예시
% \bibliography{images_and_references/main}        % main.bbl 파일 사용
%==================================================================================
%==================================================================================

%==================================================================================
%==================================================================================
%bibliography for arXiv

\providecommand{\href}[2]{#2}\begingroup\raggedright\endgroup
%==================================================================================
% \bibliographystyle{JHEP}
% \bibliography{main}
%==================================================================================
\end{document}